\documentclass[amsmath,amssymb,amsbsy,reprint,prb,preprintnumbers,showpacs,superscriptaddress]{revtex4-2}
\usepackage{graphicx,color}
\usepackage{dcolumn}
\usepackage{bm}
\usepackage{braket}
\usepackage{mathtools}
\usepackage{ulem}
\usepackage[breaklinks,colorlinks=true,linkcolor=blue,urlcolor=blue,citecolor=blue]{hyperref}
\usepackage{times}
\usepackage{physics}
\usepackage{latexsym}
\usepackage{amsmath, amssymb}
\usepackage{mathtools}
\usepackage{multirow}

\newcommand{\be}{\begin{eqnarray}}
\newcommand{\ee}{\end{eqnarray}}

\renewcommand{\theequation}{\arabic{equation}}

\begin{document}

\title{Measurement-only circuit of perturbed toric code on triangular lattice:Topological entanglement, 1-form symmetry and logical qubits}
\date{\today}
\author{Keisuke Kataoka}
\affiliation{Department of General Education, Faculty of Science and Technology, Meijo University, Nagoya 468-8502, Japan}
\author{Yoshihito Kuno}
\affiliation{Graduate School of Engineering Science, Akita University, Akita 010-8502, Japan}
\author{Takahiro Orito}
\affiliation{Department of Physics, College of Humanities and Sciences, Nihon University, Sakurajosui, Setagaya, Tokyo 156-8550, Japan}
\author{Ikuo Ichinose} 
\affiliation{Department of Applied Physics, Nagoya Institute of Technology, Nagoya 466-8555, Japan}


\begin{abstract} 
Measurement-only (quantum) circuit (MoC) gives possibility to realize the states with rich entanglements, 
topological orders and quantum memories. 
This work studies the MoC, in which the projective-measurement operators consist of stabilizers of 
the toric code and competitive local Pauli operators. 
The former correspond to terms of the toric code on a triangular lattice and the later to external
magnetic and electric fields.
We employ efficient numerical stabilizer algorithm to trace evolving states undergoing phase transitions.
We elucidate the phase diagram of the MoC system with the observables such as,
topological entanglement entropy (TEE), disorder parameters of 1-form symmetries and emergent 
logical operators.
We clarify the locations of the phase transitions through the observation of the above quantities 
and obtain precise critical exponents to examine if the observables exhibit the critical behavior 
simultaneously under the MoC and transitions belong to the same universality class.
In contrast to the TC Hamiltonian system and toric code MoC on a square lattice, the system on 
the triangular lattice is not self-dual nor bipartite, and then, coincidence by symmetries, 
such as critical behaviors across the TC and Higgs/confined phase, does not takes place.
Then, the toric code MoC on the triangular lattice provides us a suitable playground to clarify the mutual 
relationship between the TEE, spontaneous symmetry breaking of the 1-form symmetries, and
emergence of logical operators.
Obtained results indicate that toric code MoC on the triangular lattice exhibits a few distinct phase 
transitions with different location and critical exponents, and some of them are closely related with 
the two-dimensional percolation transition.
\end{abstract}


\maketitle

\section{Introduction}
Toric code (TC), which was introduce by Kitaev \cite{Kitaev_1997,kitaev2003}, has been extensively studied as a prototypical topological stabilizer code \cite{gottesman1998,RevModPhys.87.307,fujii2015}.
Its fault tolerance stems from the gauge-theoretical properties and emergent topological orders\cite{Wen_text}.
As the TC is a stabilizer system, the ground states of its Hamiltonian are obtained exactly, and 
low-energy excited states are anyons, which have nontrivial mutual braiding.
Operators creating a pair of anyons are Wilson and 't Hooft strings \cite{PhysRevD.10.2445,tHooft19781}, which play an important role in the gauge theory classifying its phase diagram.

Deformed TC model perturbed by introducing external fields is also studied extensively \cite{Trebst2008,PhysRevB.79.033109,PhysRevB.80.081104,Tupitsyn2010,PhysRevLett.106.107203,Zhang2022,Xu_2025}, in particular,
its phase diagram.
By the gauge-theoretical viewpoint, the TC state is classified as a deconfined topological state, and other gauge-theoretical states called Higgs and confinement (confined) states appear as a result of
the perturbations. 
Critical behaviors, which the system exhibits as it undergoes a phase transition, have been studied by various 
methods, including Monte-Carlo simulation \cite{Wu2012,Tupitsyn2010}, knowledge of classical spin models derived by analytical methods \cite{dennis2002,Wang2003,Timmerman2022}, etc.
The above studies are mostly performed on the TC on a square lattice, and the TC on a triangular lattice is less studied although there are a few studies \cite{Kott2024,Linsel2025}.

The previous studies are mostly based on the Hamiltonian formalism. 
As an interesting alternative, it has been shown that the TC state emerges as a steady state 
generated solely by a set of projective measurements on a quantum circuit, 
namely measurement-only circuit (MoC) \cite{Lavasani2021,PhysRevB.110.245102}. 
If one chooses a suitable combination of the measurement operators and suitably applies them to 
a many-body system sequentially as a circuit, the emergent state exhibits topological order 
similar to the one in the TC. 
Furthermore, the MoC provides possibility to generate rich steady states and also states similar to 
unconventional ground states known to appear in a specific Hamiltonian. 
So far, through the MoC, various non-trivial states have been produced and investigated 
such as volume-law entangled states \cite{Ippoliti2021,Sriram2023,PhysRevResearch.6.L042063,PhysRevB.108.094104}.
However, complete understanding of the states generated by the MoC and a guiding principle to find
Hamiltonian, the ground states of which has some relation and similarity to steady states of the MoC, are still lacking.
We would like to know under what set up phase transitions in the MoC takes place, what conditions 
determine criticality of the phase transitions emerging in the MoC and how they are related to 
known models in statistical mechanics. 
To answer these issues, systematic study is required.

Along this vein, we study a MoC where the projective measurement operators are stabilizers of the TC on the triangular lattice 
and competitive local Pauli measurements. 
The formers correspond to terms of the TC Hamiltonian on triangular lattice and the later to external fields. 
We especially focus on physical properties of emergent states in the MoC
such as (i) 1-form symmetry and topological order, (ii) topological entanglement, (iii) properties as a topological qubit. 
As practical numerical methods, we employ efficient stabilizer algorithm \cite{gottesman1998,aaronson2004} for the study on all the above properties 
in the MoC.
In addition to the fact that the triangular TC is less studied, the motivations of the present study are
as follows; 
(I) this model is not self-dual nor bipartite, and therefore, locations of various phase transitions can 
differ with each other, e.g., transition from the TC to Higgs state and confined state, 
(II) relation to the bond percolation on the triangular and hexagonal 
lattices from the perspective of the 1-form symmetries,
(III) locations of the phase transitions as well as their critical exponents clarify 
the inter-relationship between the phases, 
(IV) according to these observations, we obtain important insight into physical meanings of each phase.
Furthermore, as recently it is shown that the triangular TC merges from the color code, which is one of stabilizer codes, by a partial
condensation of anyon in the color code on a hexagonal lattice \cite{Kesselring2024}, the study on the properties of the MoC producing the triangular TC is interesting and important for clarifying 
the whole phase diagram of the color code.
We also remark that the MoC, which was recently employed to study on the square lattice
TC \cite{Lavasani2021,KOI2024_TC_MoC,OKI2024_TC_MoC,Botzung_2025}, 
is quite suitable for the present work to investigate the above mentioned properties in universal way, 
in particular, topological-entanglement entropy (TEE) \cite{Kitaev2006,Levin2006}, for which precise 
measurement by other methods is not easily performed for the two-dimensional models.

In this paper, we mostly focus our attention to the topological properties of the perturbed TC,
in particular, phase transitions and critical properties related to the topological order and 1-form symmetries.
We will take a brief look at how the Higgs-confinement phase boundary looks like in the MoC,
although interesting woks have appeared recently about this issue for the deformed TC states 
\cite{Zhu2019,liu2025_1form} and decohered density-matrix state of the TC \cite{YHChen_2024}.

This paper is organized as follows.
In Sec.~II, we introduce the TC, MoC, stabilizer formalism and observables.
In particular, we list the properties of the TC, on which we focus in the present work,
and explain specific aspects of the stabilizer formalism for the model on the triangular lattice.
In Sec.~III, after displaying the overview, we show the numerical results obtained by the stabilizer formalism, 
such as the TEE and various non-local correlation functions.
From the numerical data, we identify the locations of the phase transitions indicated by the TEE, 
spontaneous-symmetry breaking (SSB) of the 1-form symmetries, etc., and observe their critical exponents.
Detailed discussion of the phase diagram of the model is given.
Sec.~IV is devoted to discussion and conclusion.
In Appendices, we explain the details of the stabilizer formalism and methods of the error estimation in the 
finite-size scaling (FSS), etc.
\begin{figure*}[t]
\begin{center}
\includegraphics[width=7cm]{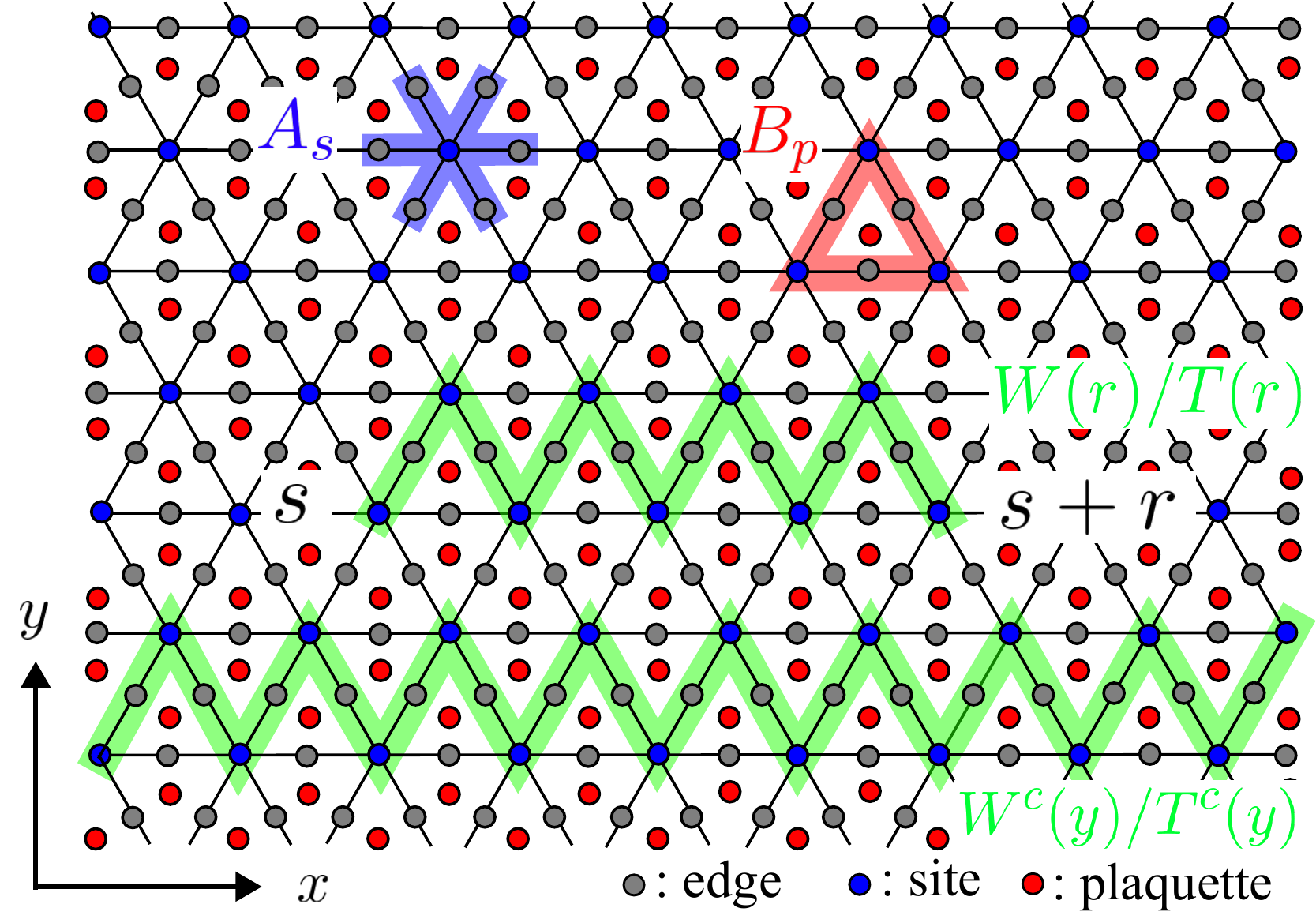}
\includegraphics[width=5cm]{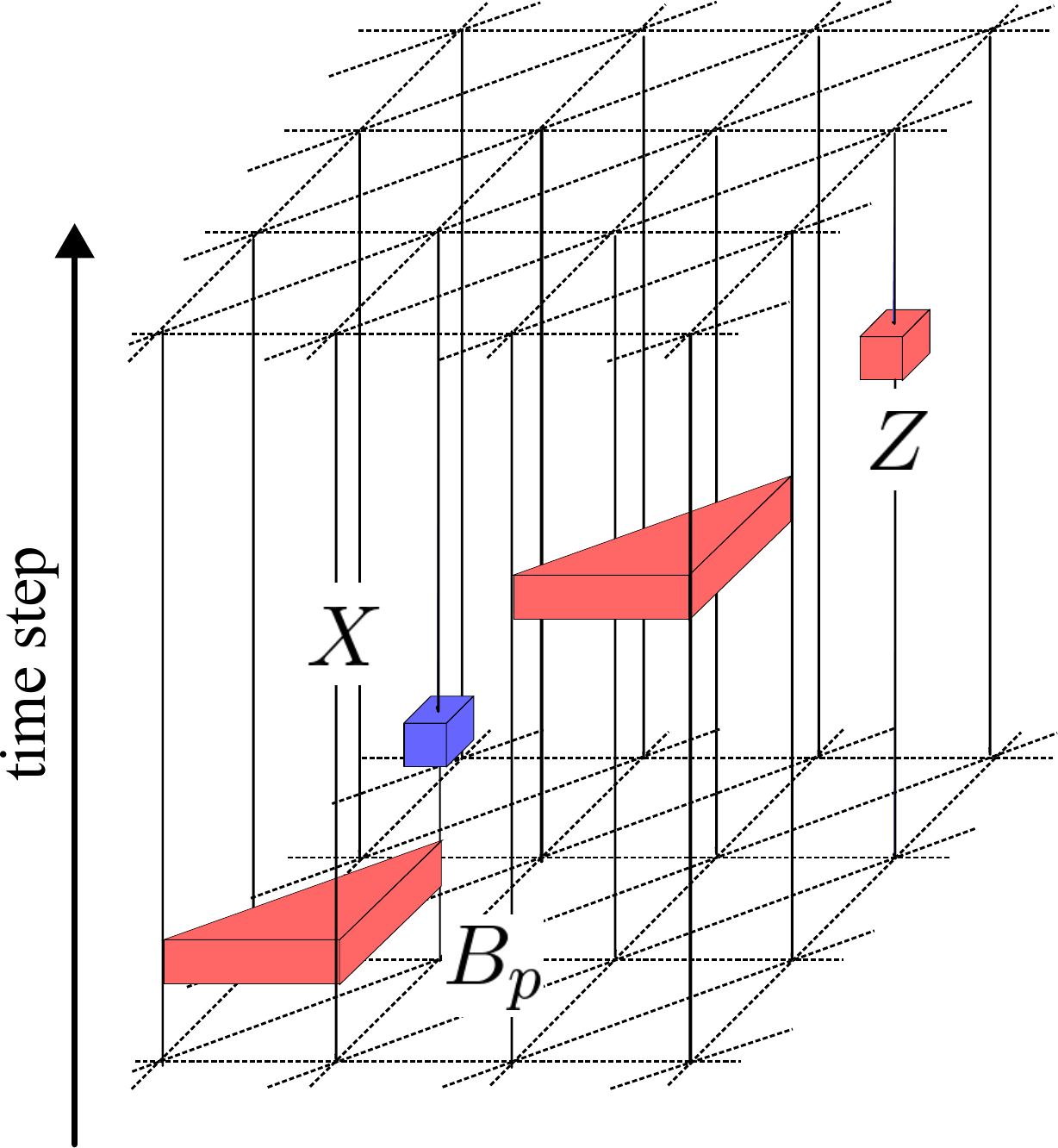}
\end{center}
\caption{Toric code on triangular lattice. 
Left panel: Stabilizers $A_s,B_p$, and loop and string, on which observables are defined.
Right panel: Schematic picture of MoC. 
Operators $X_e$, $Z_e$, $A_s$ and $B_p$ are applied to the state under consideration.}
\label{Fig1}
\end{figure*}
\section{Measurement-only circuit of toric code on triangular lattice}

In this section, we explain the set up of the MoC for the TC on the triangular lattice; 
(i) Hamiltonian corresponding to the MoC and symmetries of the target system,
(ii) Stabilizer formalism, and (iii) Observables.
As we explained in introduction, the TC system has various interesting and important aspects,
which are closely related with each other, although clear understanding of their inter-relation is still lacking.
Detailed numerical study in this work sheds light on relation between the above mentioned properties of the TC.

\subsection{Toric code system and symmetries}
To understand the target TC on the triangular lattice, the corresponding ``parent Hamiltonian" is quite 
useful, which is given as follows in terms of Pauli matrices $X_e$ and $Z_e$ on links $\{e\}$,
\be
H_{\rm TC}=-\sum_s A_s -\sum_p B_p - h_x\sum_e X_e-h_z\sum_e Z_e,
\label{HTC1}
\ee
where $A_s= \prod_{e\in s} X_e$, $e\in s$ denotes edges emanating from site $s$, 
and $B_p=\prod_{e\in p} Z_e$, $e\in p$ denotes edges surrounding plaquette $p$ (see Fig.\ref{Fig1}). 
External fields are added to the Hamiltonian with coefficients $h_x$ and $h_z$.
For the TC on the square lattice, phase diagram of the system with the external fields has been studied by various methods and is clarified.
The ground state (GS) of the TC, $|\rm TCGS\rangle$, on the square lattice for $h_x=h_z=0$ is given as follows in terms of
$A_s$ and $B_p$ defined on the square lattice similarly to the above
$A_s|{\rm TCGS}\rangle=B_p|{\rm TCGS}\rangle=|{\rm TCGS}\rangle$, and as increasing $h_x$ and/or $h_z$,
phase transition to a trivial state takes place.
We expect that the TC on the triangular lattice exhibits a similar phase diagram, and 
call the $h_x$-dominated region confinement regime and $h_z$-dominated Higgs regime 
from view point of gauge theory.
Similarly, we call the TC state and TC phase simply TC, as the standard nomenclature.

For the system given by Hamiltonian (\ref{HTC1}), the GS in the small $h_x$ and $h_z$ regime has topological orders, which are characterized as follows; 
\begin{enumerate}
\item The GSs have topological property by long-range entanglement measured by, e.g., TEE, 
which is defined in the following subsection.
\item The GSs exhibits SSB of electric and magnetic 1-form symmetries, symmetry operators of which are given by
 Wilson and 't Hooft loops. 
The topological order stems from the SSB of the 1-form symmetries \cite{Zohar2009,NUSSINOV_2009,Gaiotto_2015,mcgreevy2023}.
In the present study of the system on the triangular lattice, both the Wilson and 't Hooft loops are defined 
on the simplest zigzag loop; $\prod_{e\in\cal C}Z_e$ and $\prod_{e\in\cal C}X_e$, as it is easily seen that the loop ${\cal C}$ 
can be regarded as a closed loop on the dual honeycomb lattice as well as that on the direct triangular lattice.
See Fig.1.
We also consider open string operators on a zigzag string, which are nothing but disordered operators
of the 1-form symmetries.
These operators work as a diagnosis of anyon condensation.
\item The non-contractible Wilson and 't Hooft loops play a role of logical operaors.
The GSs are discriminated by eigenvalues of two out of four non-contractible loop operators that are 
commutative with each other.
These loops are deformable by using $A_s$ and $B_p$.
As a result, any local operators cannot distinguish the four-fold degenerate GSs.
\end{enumerate}

It is widely believed that the above three kinds of nature of the TC are closely related with each other.
For example, the SSB of the 1-form symmetry is the origin of the topological order from the viewpoint of
Landau-Ginzburg paradigm and it generates TEE.
In order to investigate the inter-relationship, it is useful to study locations and properties of the phase
transitions emerging under the MoC by observing order parameters.
As in the TC on the square lattice, for the present TC model, it is expected that a phase transition to 
the trivial states takes place as $h_x$ and /or $h_z$ are increased.
Contrary to the square lattice TC, the triangular system does not have self-duality, and therefore,
symmetry-origin coincidence of locations of phase transitions disappear, e.g., Wilson string resides on the triangular
lattice whereas 't Hooft string on the honeycomb lattice.
The condensation of these loops may emerge at different values of $h_x$ and $h_z$ \cite{Linsel2025}.

In this work, we investigate the TC system by using a specific efficient formalism, namely, the MoC.
There, the random projective measurements are applied to the target state, the operators of which are 
given by the terms in the Hamiltonian of the TC on the triangular, Eq.~(\ref{HTC1}). 
We study whether this MoC by itself produces phases related to the TC, 
and how the resulting phases acquire properties similar to those observed in the Hamiltonian formulation.
Practically, this MoC can be performed faithfully by the stabilizer
formalism~\cite{gottesman1998,Nielsen2011}, which provides us efficient numerical 
methods \cite{aaronson2004} to investigate the TC on the triangular lattice, 
schematics of which are shown in Fig.\ref{Fig1}.
For example, calculation of the TEE is very difficult by the Monte-Carlo simulation, whereas it can be
done with great accuracy\cite{Fattal2004} through the MoC as shown in this work.
In the following subsection, we briefly explain the stabilizer formalism.


\subsection{Measurement-only circuit by stabilizer formalism and observables}

The stabilizer formalism is defined by a set of Pauli operators called
stabilizer group, all of which are commutative with each other. 
For a finite lattice system, the stabilizer group has a set of finite generators described by $\{g_\ell\}$, 
where $g_\ell$'s are commutative and independent with each other.
Each operator $g_\ell$ works as a ``base'', and thus any element of the stabilizer group is generated
by multiplication of the generators.
Target pure or mixed state, $\rho$, is characterized by $g_\ell \rho=\rho$ for all $\ell$, and therefore, 
$\rho=\prod_\ell (I+g_\ell)/2$, where $I$ is the identity operator of the Hilbert space.
A trajectory of the MoC is described by a history of the set of generators of stabilizer group \cite{aaronson2004}.
In the present work, we consider the TC on the triangular lattice of the system size $(L_x,L_y)$
with periodic boundary conditions (see, Fig.1), and therefore, there are $(L_xL_y)$ $A_s$ operators
and $(2L_xL_y)$ $B_p$ operators.
By periodic boundary conditions, not all the operators are independent because of the identities such as
$\prod_{s\in \mbox{all}}A_s=1$ and $\prod_{p\in \mbox{all}}B_p=1$, and therefore, two additional constraints
have to be imposed to identify a GS of the TC as a pure state.
These constraints are given by the logical operators, and in this study we employ
$W_{i}\equiv \prod_{C_{i}}Z_\ell$ and $T_{i}\equiv\prod_{C_{i}}X_\ell$, where $C_{i} (i=x, y)$ 
are two kinds of non-contractible zigzag loops in the $x$ and $y$ directions as shown in Fig.\ref{Fig1}.
\begin{figure}[t]
\vspace{-0.5cm}
\begin{center}
\includegraphics[width=8.5cm]{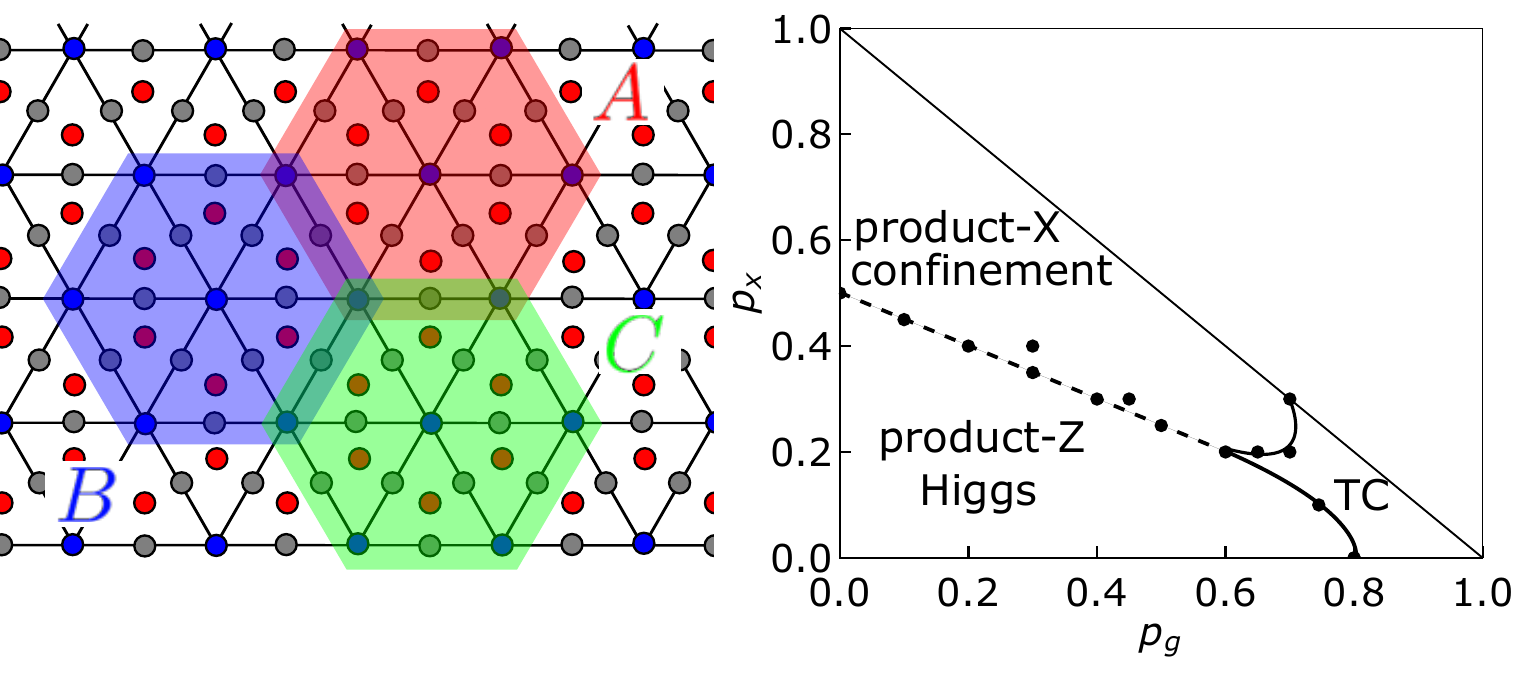}
\end{center}
\vspace{-0.5cm}
\caption{Left: Subsystems $A,B,C$ for calculation of the topological entanglement entropy (TEE).
Right: Phase diagram obtained by the TEE calculated by the MoC.
The boundary between Higgs and confinement regimes is determined by behavior of the TEE, string and loop operators, 
some of which are shown in Appendix D.
The dots are locations of the measurement.
The label ``TC'' denotes the toric code phase, a topological ordered phase. 
Data in Appendix D indicate that a Higgs-confinement phase transition takes place for $p_g=0.6$,
whereas for $p_g\leq 0.4$, at most crossover.
}
\label{Fig2}
\end{figure}
\begin{figure*}[t]
\includegraphics[width=8cm]{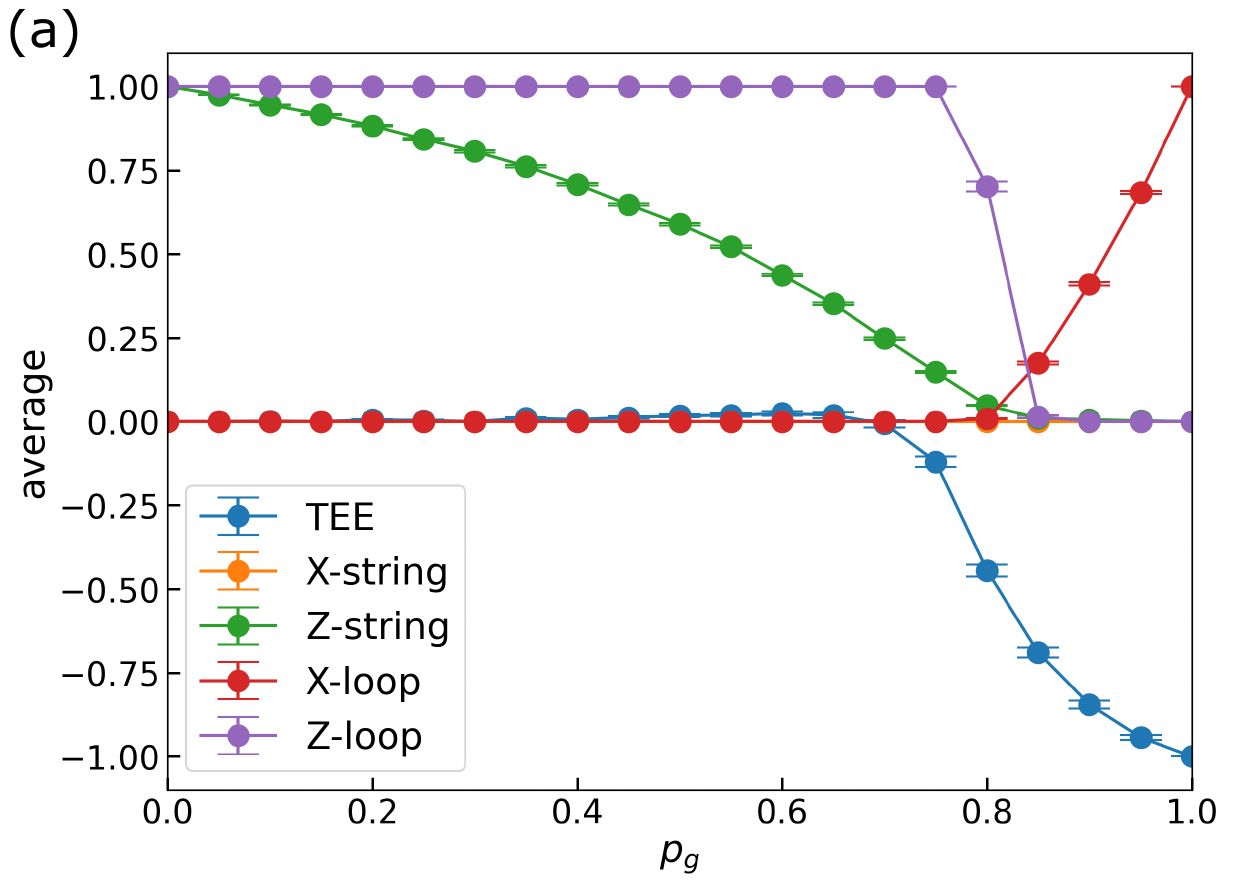}
\includegraphics[width=8cm]{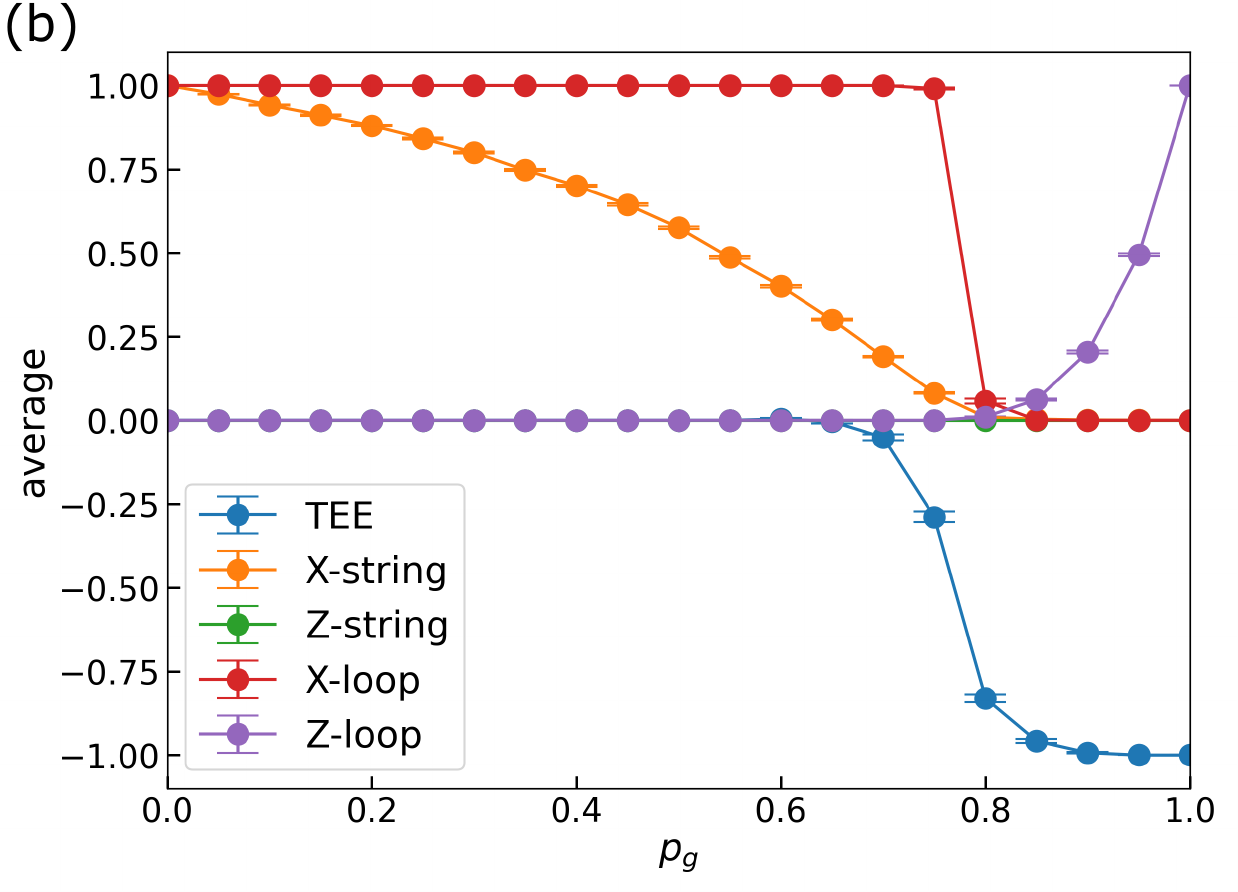}
\caption{Overview of observables from Higgs-confinement phase to TC.
There are three kinds of observables, which we investigate in detail to clarify the phase diagram and 
critical behavior of the system.
These three are; (i) topological entanglement entropy (TEE),
(ii) $X$ and $Z$ non-contractible loops ($X/Z$-loop), (iii) $X$ and $Z$ open string ($X/Z$-string).
(a) MoC starts with the $X$-product state and observed stabilizers are $Z_e$ (probability $p_z$)
and $A_s$ and $B_p$ (probability ${p_g \over 2}$ for each).
Under the MoC, $p_z+p_g=1$, whereas $p_x=0$.
The system undergoes a phase transition from the Higgs regime to the TC.
(b) MoC starts with the $Z$-product state and observed stabilizers are $X_e$ (probability $p_x$)
and $A_s$ and $B_p$ (probability ${p_g \over 2}$ for each).
Under the MoC, $p_x+p_g=1$, whereas $p_z=0$.
The system undergoes a phase transition from the confinement regime to the TC.
Detailed study reveals that the three observables exhibit a transition-like behavior at different values of
$p_g$.
}
\label{Fig3}
\end{figure*}

Target physical observables are calculated by the stabilizer formalism.
To calculate the TEE, we employ specific types of subsystems as shown in Fig.\ref{Fig2} left.
All subsystems are composed of hexagons and therefore, they have only obtuse angles with the boundary. 
(The technical details of the calculation of the TEE are briefly explained in Appendix A.)
For this partition, we calculate entanglement entropy (EE). The EE of subsystem $A$, $S_A$, is given 
by the number of stabilizers located on the boundary of $A$ \cite{Fattal2004,Nahum2017}.
 It is known and verified for the case under study that for the TC with topological order, $S_A= -L_A +\gamma$, 
where $L_A$ is the length of the boundary and $\gamma$ is a universal number 
$\gamma=-\log_2 2=-1$ \cite{Wen_text} for any composite of hexagons for $A$,
whereas $\gamma=0$ for trivial product states.
For the systems stabilized by solely $\{Z_e,X_e\}$, there are no stabilizers connecting 
subsystem $A$ and its complement $A^c$, and then, $\gamma=0$ indicating no phase transitions between these states.
The TEE is given by the combination of the EEs calculated for specific regions $A,B$ and $C$ 
(see Fig~.2) such as \cite{Kitaev2006,Levin2006},
\be
S_{\cal T}= S_A+S_B+S_C-S_{AB}-S_{BC}-S_{CA}+S_{ABC},
\label{ST}
\ee
and it is easily verified that the boundary dependent terms are canceled out with each other, and
$S_{\cal T}=\gamma$.
In the vicinity of a topological phase transition, $S_{\cal T}$ is to exhibit the step-function like behavior. 
In the MoC, we calculate the average value of $S_{\cal T}$ obtained from many trajectory samples, and therefore, the averaged $S_{\cal T}$ takes fractional values.

In addition to the TEE, we calculate the disorder parameters of the 1-form symmetries mentioned in the
previous subsection and observe the behavior of the logical qubits of states evolving in the MoC.
As shown in the previous works \cite{KOI2024}, 
the R\'{e}nyi-2 correlation functions of these operators are quite useful for the study,
which are given as follows,
\be
C_O= \rm Tr[\rho O \rho O],
\label{observe1}
\ee
where $O$ is the target operators such as $\prod_{e \in C_{ij}} Z_e$ with an open zigzag string 
$C_{ij}$ connecting sites $i$ and $j$.
For the pure state $\rho$, it is easily verified that the above R\'{e}nyi-2 correlation functions are related with the
canonical correlation functions as
\be
C_O\propto |\Tr[\rho O]|^2.
\label{observe2}
\ee
As the density matrix is given as $\rho=\prod_\ell (I+g_\ell)/2$, $C_O$ is unity if the operator $O$ commutes 
all of the stabilizers $\{g_\ell\}$, otherwise zero. 
Then, the evaluation of the correlation functions can be performed quite accurately.
As explained in the above, the correlation functions of the string operators are the disorder parameters 
of the target 1-form symmetries, which can be evaluated more easily and exhibit a clearer signal of the 
phase transitions than the order parameter, i.e., the Wilson and 't Hooft closed loop operators.

\section{Numerical calculations}
In this section, we show the results obtained by the numerical calculation of the observables;
the TEE, disorder parameters of the 1-form symmetries and emergence/disappearance of the logical
operators.
Detailed study of the above observables provides us locations and critical exponents of the phase 
transition.

\subsection{Overview} 
Before going into details, we display overview of typical behavior of the order parameters.
In the MoC, the operation of $X_e$ and $Z_e$-measurement corresponds to adding these into the stabilizer group, 
and removing stabilizers that are anti-commutative with them. 
This procedure makes the state under consideration enhance anyon hopping, i.e., anyon condensation.
In the practical process of the MoC, probabilities applying $X_e$ and $Z_e$ measurements are denoted 
by $p_x$ and $p_z$, respectively.
On the other hand, measurements of $A_s$ and $B_p$ are performed with equal weight for total probability $p_g$. 
We display the global phase diagram by observing the TEE as shown in Fig.~\ref{Fig2} right. 
We find that there are three types of steady states, each of which is discussed in detail later on.

Fig.~\ref{Fig3}(a) shows the numerical results for a typical MoC, in which the initial state is the product-$X$ state,
$|X_e=1\rangle^{\otimes N}$ ($N=3L_xL_y$), and then, the MoC is composed of $Z$ and $A_s/B_p$ measurements with 
the constraint $p_z+p_g=1$.
We observe states evolving for a sufficiently long period with the total measurement steps $> 10N$,
and focus on steady states. 
(See the verification of our numerics in Appendix A.)
For small $p_g$ (large $p_z$), $X$-stabilizers disappear by mostly applied $Z$-stabilizers and 
$B_p$ stabilizers, and the state evolves into a trivial product-$Z$ state as indicated by finite values of the $Z$-string and $Z$-loop operators, whereas the TEE vanishes.
On the other hand for sufficiently large $p_g$, $A_s$-stabilizers and $B_p$-stabilizers dominate, and as a result, a TC state emerges.
Interestingly enough, although single $X$-stabilizers in the initial state are eliminated by $B_p$-stabilizers,
$X$-non-contractible loop operators (logical operators) survive as they commute with $B_p$-stabilizers.
There exists a phase transition between these two states, the location of which is
roughly estimated $p_{gc}\simeq 0.8$, although the three observables seem to indicate slightly different transition points.
This discrepancy will be studied in detail in the subsequent sections.

Another example is shown in Fig.~\ref{Fig3}(b). 
The initial state is taken as the $Z$-product state and measurements of $X$, $A_s$ and $B_p$-stabilizers
are performed as in the previous case.
As $p_g$ is increased, the $Z$-product state gets losing the $Z$-strings due to the $A_s$ stabilizers, 
while keeping a finite value of the $Z$-loop as it commutes with $A_s$'s, and a phase transition to 
a TC takes place.
The emergent TC is an eigenstate of the $Z$-logical operators for this MoC contrary to the previous one. 

The results in Fig.~\ref{Fig3} clearly show a way to construct desired TC state starting from the trivial 
product states.
Furthermore, as the phase transition to the TC takes place at $p_g \sim 1$, 50\% measurement of both $A_s$ and $B_p$-stabilizers is sufficient for emergence of the TC state.

\begin{figure}[t]
\begin{center}
\includegraphics[width=6.5cm]{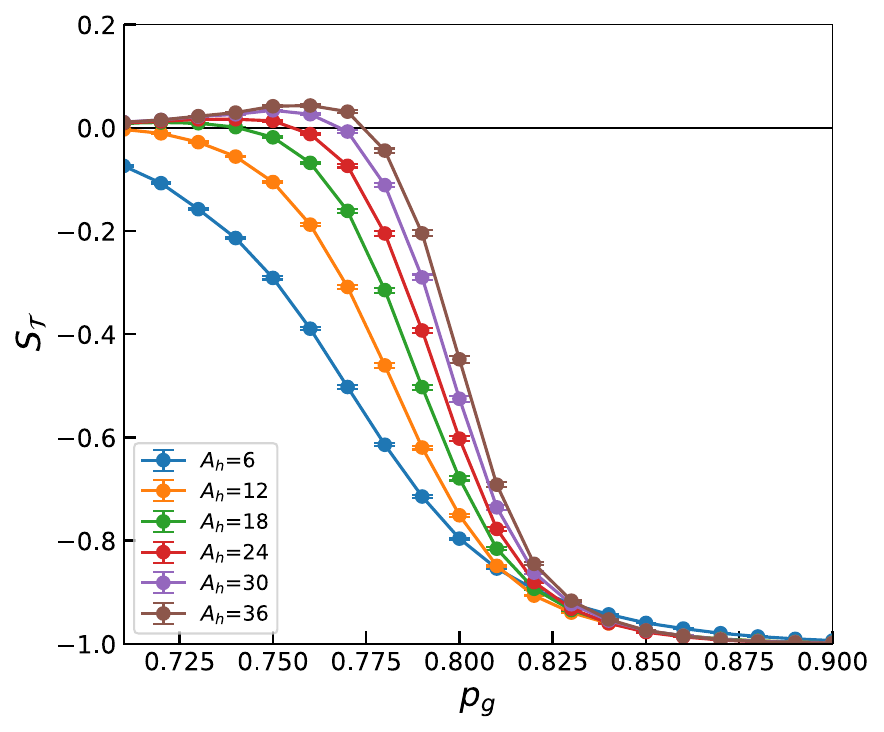}
\includegraphics[width=6.5cm]{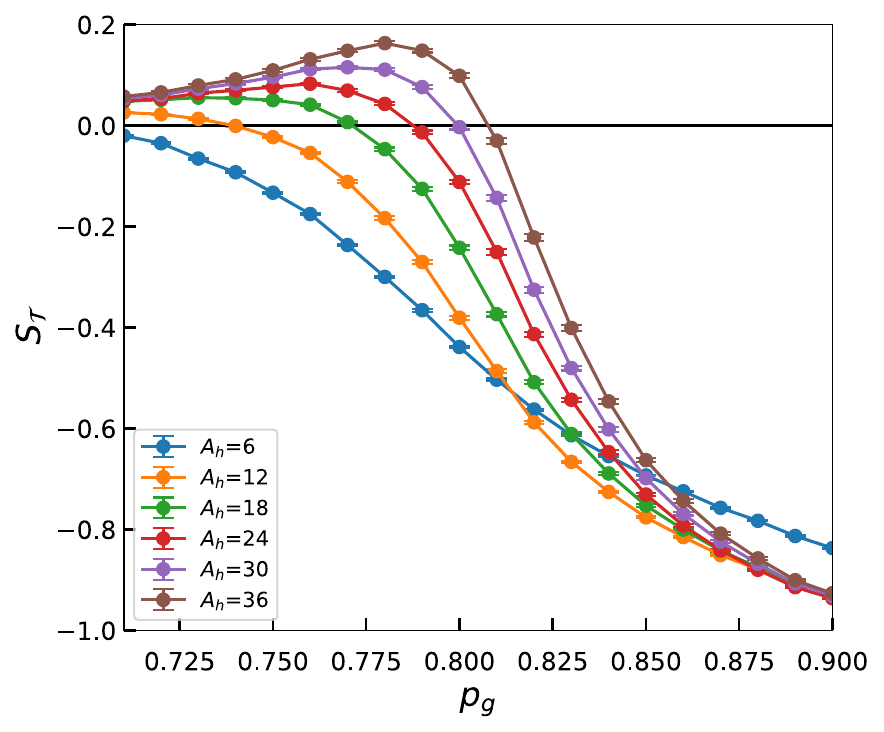}
\end{center}
\caption{Topological entanglement entropy for various system sizes.
Top: Transition from confinement regime to TC phase in the parameter space of $p_x+p_g=1$.
Bottom: Transition from Higgs regime to TC phase in the parameter space of $p_z+p_g=1$. 
$A_h$ is area of hexagon complex in unit of triangle.
The system size is $(L_x,L_y)=(24,24)$.}
\label{Fig4a}
\end{figure}
\begin{figure}[t]
\begin{center}
\includegraphics[width=7.5cm]{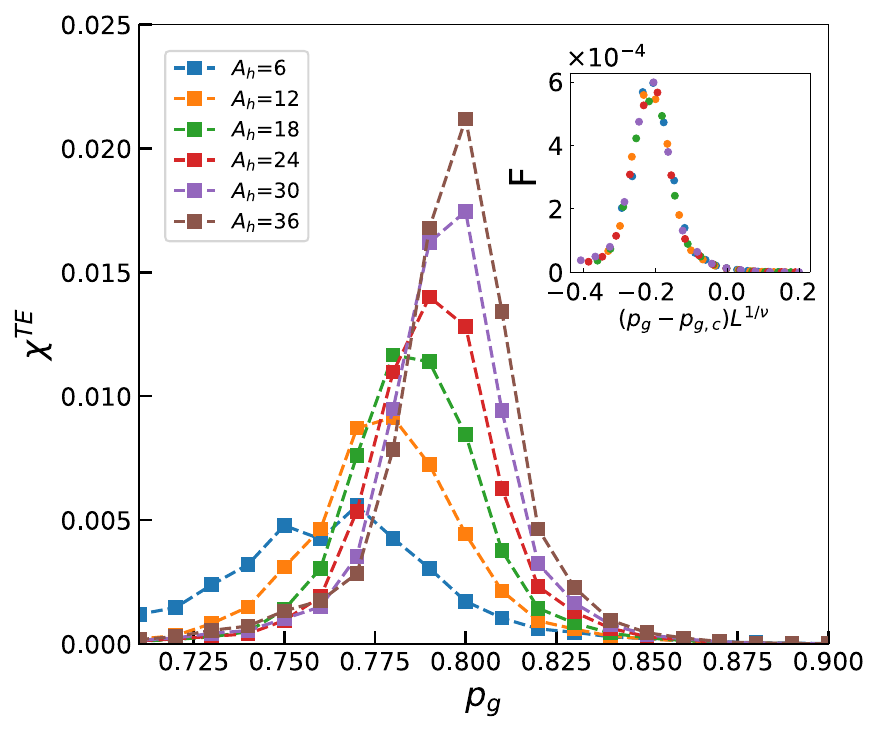}
\includegraphics[width=7.5cm]{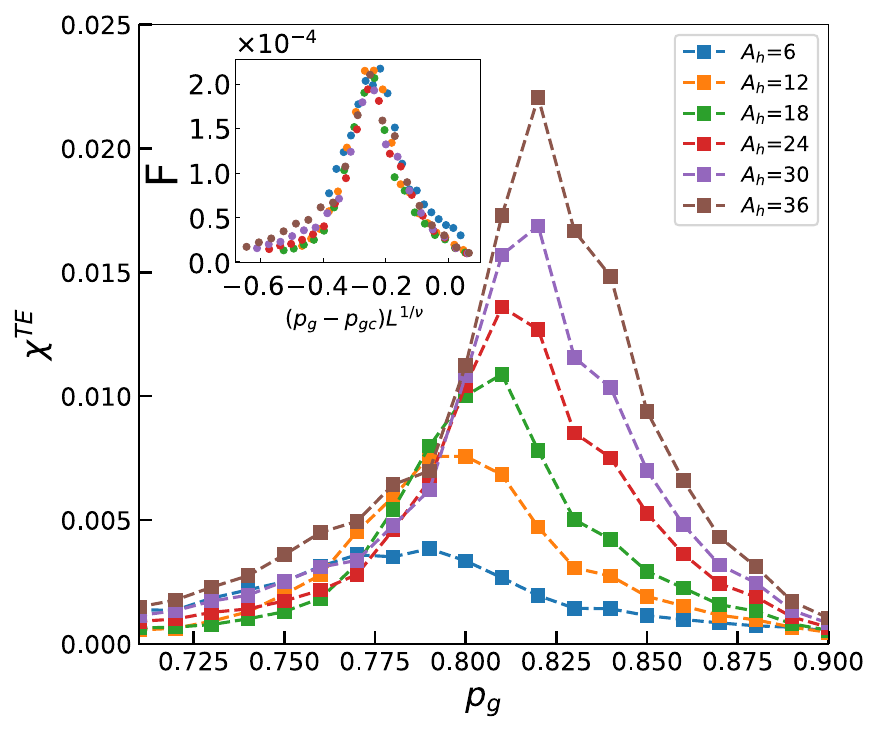}
\end{center}
\caption{Variance of topological entanglement entropy for various system sizes.
Top: Transition from confinement regime to TC phase in the parameter space of $p_x+p_g=1$.
Bottom: Transition from Higgs regime to TC phase in the parameter space of $p_z+p_g=1$.
$A_h$ is area of hexagon complex in unit of triangle.
The insets show the data collapse by the finite-size scaling.}
\label{Fig4b}
\end{figure}
\subsection{Topological entanglement entropy and phase diagram}

In order to calculate the TEE of states emerging in the MoC, we employ complexes of hexagon as subsystems. 
We found that this shape is well-suited for the calculation of the TEE by the stabilizer formalism,
as only obtuse angles appear and the numbers of $A_s$ and $B_p$ residing on both $A$ and its
compliment $\bar{A}$ can be unambiguously defined. 

We first consider the MoC, in which the initial state is the product state 
$|Z_e=+1\rangle^{\otimes N}$(or $|X_e=+1\rangle^{\otimes N}$),
and then, we apply $X_e(Z_e)$, $A_s$ and $B_p$, as we explained in the above.
Typical time step is $10 N$, which is set to satisfy the condition that the state sufficiently reaches 
 a steady state.
Figure \ref{Fig4a} displays the TEE $S_{\cal T}$ for various subsystem sizes, where as a typical system size $L$, 
we employ $L=\sqrt{A_h}$ where $A_h$ is the area of the hexagon complex.
We also observe its variance,
$\chi^{\rm TE}(p_g)$, in Fig.~\ref{Fig4b}.

The obtained data clearly show the phase transition from the $Z(X)$-product state to 
the topologically-ordered TC in which non-contractible Wilson ('t Hooft) loops have a definite eigenvalue.
(See Fig.~\ref{Fig3}.)
From the variance, we can estimate the critical value of $p_g$ and critical exponents by using the finite-size scaling (FSS) ansatz such as,
\be
 \chi^{\rm TE}(p_g)=L^\zeta F((p_g-p_{gc})L^{1/\nu}),
 \label{FSS}
 \ee
where $p_{gc}$ is the critical probability, $F$ is a scaling function, and $\zeta$ and $\nu$ are critical exponents. 
The later is a correlation-length critical exponent.
Estimated values of the critical probability and exponents for the MoC with $p_x+p_g=1$ 
starting from the $Z$-product state are (for the data collapse, see the inset in Fig.~\ref{Fig4b})
$p^x_{gc}=0.846\pm 0.016, \nu^x= 1.69\pm 0.26$.
[In this work, errors are mostly estimated by using pyfssa \cite{melchert2009,pyfssa}].

We also investigate how the state evolves under the $Z$, $A_v$ and $B_p$ measuring MoC
stating from the $X$-product state, that is, MoC with $p_z+p_g=1$.
Numerical results are show in Fig.~\ref{Fig4a} and Fig.~\ref{Fig4b}, and estimated values are 
$p^z_{gc}=0.876\pm 0.048, \nu^z= 1.74\pm 0.29$.
We observe that $p^z_{gc}$ is slightly larger that $p^x_{gc}$, whereas critical exponent $\nu$ is close
in the both cases.
We expect that the discrepancy of the critical probability comes from the fact that the Wilson loop (string)
resides on the direct triangular lattice, whereas 't Hooft loop (string) on the dual honeycomb lattice,
as well as the different shapes of $A_s$ and $B_p$.
In the two-dimensional (2D) bond percolation, the values of the threshold are quite different in
the two lattices, i.e., threshold of the honeycomb lattice is larger than that of the triangular lattice.
However, we note that two critical points in the present MoC are obviously much closer with each other than those of the 2D bond percolation.
We think that this comes from the fact that the ordinary percolation takes place in the system 
without any correlations
and its threshold is determined solely by a probability theory, whereas a state in the MoC evolves under 
the various competitive measurements as a trajectory, and therefore, the \textit{threshold} can be different 
from the value of the genuine percolation.
In fact, the stabilizers $A_s$ and $B_p$ are conjugate operators under the duality, and they are applied to 
the states in the MoC with equal probability.
This fact induces close thresholds for the two MoCs.

The result of the correlation length exponent $\nu$ indicates that the two phase transitions belong to the same universality class of $\nu\simeq 1.7$, which is apparently different from the 2D percolation with $\nu=4/3$ 
as well as the 3D percolation with $\nu=0.88$ \cite{stauffer2018}.
Then, the above observations will be compared with the numerical study on the string operators in the following subsection.
The precise universality class of the TEE transition cannot be conclusively identified.
Nevertheless, our critical exponent analysis presented in Appendix B provides us quantitative evidence that strongly constrains its possible classification.

Similar analysis to the above is performed for more general cases with non-vanishing $p_z$ and $p_x$.
We obtain a phase diagram of the present MoC by using data of the TEE for small system sizes,
which is displayed in Fig.~\ref{Fig2}.
The phase diagram is close to the expected one as we explained in Sec.~II A.

Here, we comment on the `phase' boundary of the Higgs and confinement regimes.
In the phase diagram in Fig.~\ref{Fig2}, for $p_g<0.4$, the TEE exhibits only small decrease in the region
$p_x\sim p_z$, whereas the string and loop operators, which we introduce in the following subsections, behave
quite differently in the two region and work as a good diagnosis to discriminate the two `phases'.
For $0.4<p_g\leq 0.6$, the TEE tends to exhibit a sharp decrease and a cusp at $p_x=p_z$ indicating existence of 
a sharp transition.
These numerical results of the TEE, string and loop are shown in Appendix D.

\subsection{Condensation of string operators; disorder parameters of $1$-form symmetries}
\begin{figure}[t]
\begin{center}
\includegraphics[width=8.5cm]{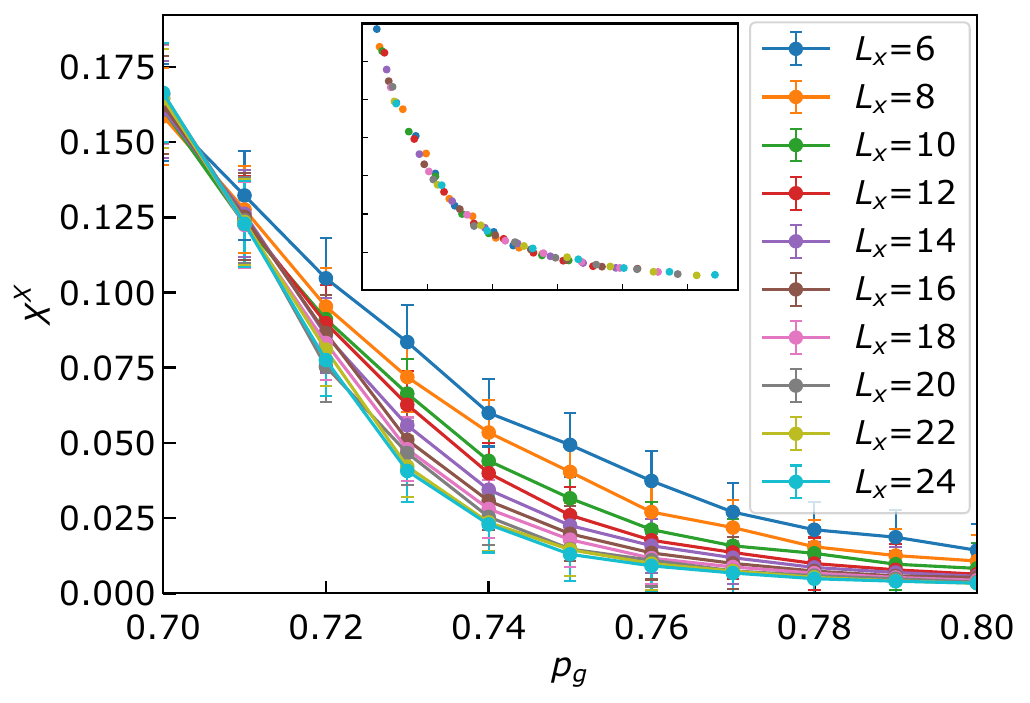}
\end{center}
\caption{Calculation of the expectation value of $X$-string from 
the confinement regime to toric code for $p_x+p_g=1$.
The decrease of $\chi^X$ comes from the increase of $\{B_p\}$ in the stabilizer group.
In the inset, we observe the collapse of the date by the finite-size scaling. 
}
\label{Fig_stringloop}
\end{figure}

In this subsection, we study the Wilson and 't Hooft strings residing on the zigzag strings in Fig.~\ref{Fig1}.
To this end, we define them for the zigzag string ${\cal L}(s,s+r)$ in Fig.~\ref{Fig1},
\be 
W(r) = \prod_{e\in {\cal L}(s, s+r)}Z_e, \; T(r) = \prod_{e\in {\cal L}(s, s+r)}X_e,
\label{stringop}
\ee
where $s$ and $s+r$ denote sites at which the zigzag string ends, and $r$ takes positive integers. 
These strings creates anyons on their edges, and measure the condensation $e$-anyon and $m$-anyon, 
respectively.
In the stabilizer formalism, the locations of a pair of anyons $(s,s+r)$ determine expectation values
of string operators for each state.
The string operators in Eq.~(\ref{stringop}) correspond to the Fredenhagen-Marcu operator (FMO) in the gauge theory \cite{Fredenhagen1983,PhysRevLett.56.223}, 
which was invented as an order parameter for the Higgs phase.
In gauge systems coupled with matters, the ordinary Wilson loop generally obeys the perimeter law, and 
it cannot discriminate the Higgs phase from Coulomb and confinement phases.
In particular, the FMO and the above corresponding observables work quite efficiently for the $Z_2$
gauge theory as we show in the present study, while the simple Wilson loop does not \cite{KOH2025_v1}.
In the stabilizer point of view, a finite value of $\langle W(r)\rangle (\langle T(r)\rangle)$ means that the both ends of the string
are included in a single void of the stabilizer $A_s (B_p)$, which is produced by 
$Z(X)$-measurements \cite{KOI2024,KOH2025_v1}.
It is obvious that this picture is closely related to the two-dimensional bond percolation. 
In Appendix C, we show a brief explanation of the FMO and its qualitative relationship to our target string 
(or loop) 
order parameters by showing the behavior of these diagnoses for the Higgs-TC transition.
Although the FMO is a widely accepted diagnosis, we employ other order parameters, 
which are more tractable in the stabilizer formalism.

\begin{figure*}
\begin{center}
\hspace{-1cm}
\includegraphics[width=18cm]{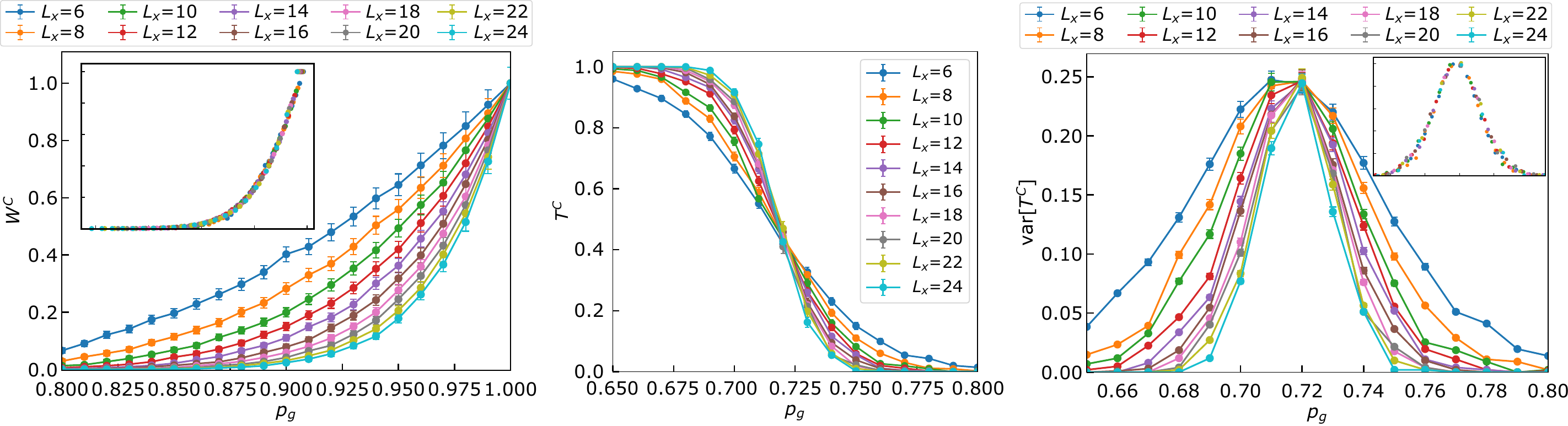}
\end{center}
\caption{
Order parameters as a function of $p_g$ on the line $p_x+p_g=1$.
Left panel: Calculation of the expectation value of $Z$-loop from the 
confinement regime to toric code.
The increase of $W^c$ comes from the decrease of $\{X_e\}$ in the stabilizer group.
In the inset, we observe the collapse of the date by finite-size scaling.
Center panel: Calculation of the expectation value of $X$-loop from 
the confinement regime to toric code.
The decrease of $T^c$ comes from the decrease of $\{X_e\}$ in the stabilizer group.
Right panel: Variance of $X$-loop, and the inset shows collapse of data by FSS. 
}
\label{Fig_loop}
\end{figure*}

Correlation functions of the string and their `susceptibility' observing the long-distance behavior
of the state are defined as,
\begin{eqnarray}
C^Z(r,y) &=& \mbox{Tr} [\rho W(r) \rho W(r)], \nonumber\\ 
C^X (r,y)&=& \mbox{Tr}[ \rho T(r) \rho T(r)],
\label{correlationF}
\end{eqnarray}
\begin{eqnarray}
&\chi^Z = \frac{1}{(L_x-1)L_y} \sum_{y=1}^{L_y} \sum_{r=1}^{L_x-1}C^Z(r,y), \nonumber \\
&\chi^X = \frac{1}{(L_x-1)L_y} \sum_{y=1}^{L_y}\sum_{r=1}^{L_x-1}C^X(r,y),
\label{chi}
\end{eqnarray}
where $y$ denotes $y$-coordinate of the zigzag string, and
we take the $L_y$-ensemble average in the $y$-direction of the lattice. 
As we mentioned above, $C^Z(r,y) = |\langle W(r)\rangle|^2$, etc for we are considering pure states. 
In the following practical investigation with the MoC, we calculate the mean values of 
the observables using the above correlation functions and their susceptibility obtained for
many trajectory samples.

Figure \ref{Fig_stringloop} displays numerically obtained results.
Under the MoC with $p_z=0$ and $p_x+p_g=1$, $\chi^{X}$ exhibit a critical behavior from the confinement regime
to the TC, in which the other disorder parameter $\chi^Z$ vanishes. 
The decrease of $\chi^X$ comes from the increase of $\{B_p\}$ in the stabilizer group.
We notice that the behavior of $\chi^X$ as a function of $p_g$ is very close to that of magnetization in a finite system 
as a function of temperature, i.e., it decreases smoothly as $p_g$ increases and vanishes for $p_g > 0.8$.
(See also Fig.~\ref{Fig_Astringloop} in Appendix D.)
This behavior is consistent with the expectation that the 1-form symmetries are spontaneously broken 
in the TC and the string operators work as disorder parameters of the 1-form symmetries.
Furthermore, $\chi^X$ exhibits only small system-size dependence. 
However, we find that the FSS is well applied to $\chi^X$ and a good scaling collapse appears with the critical
value and exponent such as (see the inset)
$p^x_{c,{\rm st}}=0.723\pm 0.013,\nu^x_{\rm st}=1.63\pm0.35$.
These values should be compared with those of the non-contractible loop, which are studied in the
following subsection.
This observation is strongly connected with the relationship between the 1-form symmetries and emergence
of logical qubits.

Contrary to the 't Hooft string, we observe that the Wilson string is vanishingly small in the MoC process from
the confinement to TC as seen in Fig.~\ref{Fig3}(a).
This result is plausible since the 1-form symmetry with respect to the symmetry operator 
$\prod_{e\in {\cal C}}Z_e$ (${\cal C}$ is an arbitrary closed loop) is not realized in both the confinement (destroyed by the dominant $X$-measurement) and TC phase (SSB).

In Appendix D, we show the numerical data for the MoC with $p_z+p_g=1$, 
and the estimated values are $p^z_{c,{\rm st}}=0.787\pm 0.012,\nu^z_{\rm st}=1.51\pm0.31$, respectively.

\subsection{Critical properties of emerging logical operators}

In the previous subsection, we observed the critical behavior of the $Z(X)$-string as the system undergoes
the transition from the Higgs (confined) state to TC.
In this subsection, we investigate the logical $X(Z)$-operator in the present MoC, which is a hallmark of the TC,
that is, the existence of the logical operators is a diagnosis of the emergence of the logical code in the TC. 

In this study, we focus on 't Hooft (Wilson) loops residing on the non-contractible zigzag loops
winding in the $x$-direction (see Fig.~\ref{Fig1}), and define the logical operators as,
\be 
W^c(y)=\prod_{e\in C(y)}Z_e, \; T^c(y)=\prod_{e\in C(y)}X_e,
\label{logOp}
\ee
where $C(y)$ $(y=1,\cdots, L_y)$ denotes non-contractible $y$-th zigzag loop in the $x$-direction, 
and all data are averaged over $L_y$ loops; $W^c\equiv \frac{1}{L_{y}}\sum_y\langle W^c(y)\rangle, T^c\equiv \frac{1} {L_{y}}\langle\sum_yT^c(y)\rangle$.
Values of $W^c$ and $T^c$ depend on the density of the stabilizers $\{X_e\}$ and $\{Z_e\}$
and independent of that of $\{A_s\}$ and $\{B_p\}$.
 
We show the data for typical MoC starting from the $Z$-product state and applying $X$, $A_v$ and $B_p$
measurements with the constraint $p_x+p_g=1$.
For $p_x\gg p_g (p_g\gg p_x)$, the system is in the confinement (TC) regime, and there exists a transition
between the two regimes as we observed in the previous subsection.
Here, we examine it from the viewpoint of the logical qubit.

We first observe the behavior of the Wilson loop from the confinement regime to the TC.
Numerical data displayed in Fig.~\ref{Fig_loop} , $W^c$, starts to increase as the state transforms into 
the TC phase, indicating that the emergent TC is an eigenstate of the logical operators
$\prod_{e\in {\cal C}_i}Z_e$, where ${\cal C}_i (i=x,y)$ are non-contractible loops in $x(y)$ direction.
This is a result of the choice of the initial state, that is, edge $Z$-stabilizers merge with each other 
by applied $A_s$-stabilizers to form non-contractible operators, although most of them 
disappear by the measurement of $X$-stabilizers.
FSS can be applied suitable for $W^c$, and it provides us 
$p^z_{c,{ \rm loop}}=1.009\pm 0.017,\nu^z_{\rm loop}=1.39\pm 0.27$.

Next, we study the behavior of the non-contractible 't Hooft loop, $T^c$, from the confinement regime to the TC.
Numerical results are shown in Fig.~\ref{Fig_loop}.
$T^c$ exhibits a step-function like behavior in the critical regime with system-size dependence.
In order to extract the transition point and critical exponent, we calculate the variance of $T^c$
to find that a sharp peak emerges.
Although, height of peaks do not develop for increasing system size, the peak is getting sharper.
FSS gives us the values such as, 
$p^x_{c,{ \rm loop}}=0.719\pm 0.003,\nu^x_{\rm loop}=1.27\pm 0.06$.

For the transition from the confinement to the TC phases, it is interesting and important to observe that 
the $X$-string, which is the disorder parameter of the 1-form $X$ symmetry, vanishes first and then, 
the $Z$-loop starts to get a finite value as our numerical data reveal.
This behavior supports the general belief that the 1-form-symmetry is SSB in the TC.
We also notice that the values of the critical exponent $\nu$ of all the string and loop order parameters are
close to that of the 2D percolation value, $4/3$.
This is a sharp contrast to the TEE.
More discussion on this point will be given in Sec.~IV.

Calculations of the other case  $p_z+p_g=1$ are shown in Appendix D, which exhibit similar behaviors.
 
Here, we comment that the above critical behavior of 1-form symmetry breaking and also emergence of the logical operators may be specific in the present MoC, in other words, this finding may not be universal.
Another set up for generating states and another numerical method, such as the Monte-Carlo simulation, 
can reveal different critical behavior of the system under the transition from the Higgs/confined regime 
to the TC.
However, we think that the findings of this study shed light on the structure of the Hamiltonian system
of the triangular TC.


\section{Conclusion and discussion} 

In this work, we studied the MoC of the perturbed TC on the triangular lattice, and investigated its phase diagram and critical behavior.
The observables that we employed are the TEE and non-local correlation functions such as the Wilson/'t Hooft
loop and string, which play an important role for diagnosing topological properties of the system with 1-form symmetries.
We found that the critical behaviors of all the above observables take place in a small parameter region,
and therefore, we observe that physics observed by the order parameters are closely related with each other,
supporting the common belief.
As the target system is defined on the less-symmetric triangular lattice compared with the square lattice, 
this observation is rather conclusive.
 
We first study the TEE and identified the location of the phase transition from the Higgs/confinement to 
the TC states to find a small but finite discrepancy of the duality explicitly broken by the shape of the spatial 
lattice (the triangular lattice), whereas it is small from the view point of the percolation.
In fact, the critical values of $p_g$ are fairly close to that obtained for the perturbed TC on the square lattice.
We think that this indicates the fact that the TEE is given by the numbers of stabilizers
on the boundaries between subsystems and by canceling out the contributions depending on the 
length of the boundaries [Eq.~(\ref{ST})], and therefore, the TEE does not strongly depends on 
the geometry of the lattice.

\begin{table*}[htbp]
\centering
    \begin{tabular}{|c|c|c|}    
    \multicolumn{3}{c}{Table 1A}    \\\hline
    \multicolumn{3}{|c|}{$p_x+p_g=1$}     \\ \hline
    \multicolumn{1}{|c|}{observable}   &  $p_g$  & \multicolumn{1}{c|}{$\nu$}  \\ \hline
    TEE$^\ast$   
    & 0.846 $\pm$ 0.016& 1.69 $\pm$ 0.26\\
    $X$-string& 0.723 $\pm$ 0.013& 1.63 $\pm$ 0.35\\
    $Z$-loop 
    & 1.009 $\pm$ 0.017& 1.39 $\pm$ 0.27\\
    $X$-loop  
    & 0.719 $\pm$ 0.003
    & 1.27 $\pm$ 0.06   \\ \hline
  \end{tabular}
   \begin{tabular}{|c|c|c|}    
    \multicolumn{3}{c}{Table 1B}    \\\hline
    \multicolumn{3}{|c|}{$p_z+p_g=1$}     \\ \hline
    \multicolumn{1}{|c|}{observable}   &  $p_g$  & \multicolumn{1}{c|}{$\nu$}  \\ \hline
    TEE$^\ast$   & 0.876 $\pm$ 0.048& 1.74 $\pm$ 0.29\\
    $Z$-string& 0.787 $\pm$ 0.012  & 1.51 $\pm$ 0.31\\
    $Z$-loop & 0.786 $\pm$ 0.002 & 1.45 $\pm$ 0.14\\
    $X$-loop  & 0.856 $\pm$ 0.034 & 1.37 $\pm$ 0.17\\ \hline
  \end{tabular}
\vspace{0.5 cm}
\caption{Table 1A: Values of critical point and critical exponent obtained from the TEE, $Z$-string, and 
$Z(X)$-non-contractible-loops for the MoC with keeping $p_x+p_g=1$.
Table 1B:Values of critical point and critical exponent obtained from the TEE, $Z$-string, and 
$Z(X)$-non-contractible-loops for the MoC with keeping 
$p_z+p_g=1$.
TEE* is the value obtained by the bootstrap method. (See Appendix B.)}
\label{Tab1}
\end{table*}

We found that the critical exponent of the TEE correlation length differs from the corresponding 
quantity in the percolation.
We also observed that the critical exponents of the string and particularly the loop correlations
are close to that of the 2D percolation as in the TC on the square lattice \cite{Botzung_2025}.
These findings imply that the entanglement of quantum system measured by using subsystems
may not be directly related to the percolation, which reflects a global `stringy' property of the system.
To our best knowledge, this issue has not be addressed yet, although the TEE is a good diagnostic
quantity of topological phase.
Further study is required to get deep understanding of the \textit{TEE in the critical regime}.
We remain it as a future problem.

After mapping out the phase diagram by means of the TEE, we studied the non-local correlation functions 
in the MoC.
As expected, the behavior of the string operators as a function of $p_g$ is consistent with the fact that they are disorder parameters of the 1-form symmetries, that is, the $Z(X)$-string operators has a finite 
expectation value in the Higgs/confinement state and vanishes in the TC.
We located the critical point by using the FSS and found that the critical value of $p_g$ for the SSB of the 1-form symmetries is slightly smaller than that of the TEE, whereas
the critical exponent $\nu$ is close to that of the 2D percolation. (See Table~\ref{Tab1}.) 
This result is interesting and important as it supports the observation that the \textit{TEE phase transition} 
does not coincide with the \textit{emergence of the SSB of the 1-form symmetries,} whereas it is widely believed that
the SSB is the origin of the topological order.
More precisely, the SSB takes place first, and then, the TEE gets a finite value \cite{note1}. 
As mentioned in the preceding paragraph, this discrepancy may come from the global stringy property of 
the 1-form symmetries and their global deformability, whereas the TEE is evaluated by the numbers of
stabilizers on boundaries connecting adjoining subsystems.
More precisely, a finite value of strings $\langle W(r)\rangle (\langle T(r)\rangle)$ comes from the states in which the both ends 
of the string are included in a single void of the stabilizer $A_s$$(B_p)$
produced by $Z(X)$-measurements \cite{KOI2024,KOH2025_v1}.
This condition is obviously related to the percolation and is different from the condition of the finite 
TEE, although a set of stabilizers play an essential role for both of them.

As far as we know, this kind of discrepancy in behaviors of the order parameters of the TC in the critical
regime has been rarely noticed and discussed.
Observation about somewhat related discrepancy has been given for a very specific deformed TC model 
that is exactly solvable~\cite{PhysRevB.77.054433,10.21468/SciPostPhys.15.6.253}.
In this work, we remain the issue how the TEE precisely relates to the global stringy properties
of the system, in particular, the 1-form symmetries as a future problem,
although the TEE measures the stability of the loop-gas picture of the TC.

Finally, we studied the expectation value of the logical qubits, the non-contractible Wilson and 't Hooft loops.
We found that they exhibit clear critical behavior and they acquire a finite value as $p_g$ increases.
We estimated the transition points and the correlation critical exponent by the FSS.

We here summarize the transition points and criticalities in Table \ref{Tab1}.
As we explained in the Sec.~III, the MoC with keeping $p_z=0$ is more stable than that of $p_x=0$, and this 
behavior comes from the fact that $Z$-measurement is competitive with $A_s$ and disturbs states
more strongly than $X$-measurement.
We also found that the non-contractible loop operators exhibit similar behaviors of the corresponding
string operators, and reflect the 2D percolation property.

As stated previously, the findings of this study may be specific ones through the MoC.
Nonetheless, the present study employing the MoC sheds light on the phase structure of 
the deformed TC Hamiltonian on the triangular lattice, and the obtained observation in this work 
will help for getting deep understanding of topological phase transitions.
We hope that our work opens several significant questions and research directions.

\section*{Data availability}
The data that support the findings of this study are available from the authors upon reasonable request.\\

\bigskip
{\it Acknowledgements.---}
This work is supported by JSPS KAKENHI: JP23K13026(Y.K.) and JP23KJ0360(T.O.). The computations in
this work were done using the facilities of the Supercomputer Center, The Institute for Solid State Physics,
The University of Tokyo.

\bibliography{ref2}

\renewcommand{\thesection}{A\arabic{section}} 
\renewcommand{\theequation}{A\arabic{equation}}
\renewcommand{\thefigure}{A\arabic{figure}}
\setcounter{equation}{0}
\setcounter{figure}{0}
\appendix

\setcounter{section}{0}
\renewcommand{\thesection}{\Alph{section}}
\makeatletter
\renewcommand{\theHsection}{\Alph{section}}
\makeatother



\makeatletter
\renewcommand{\theHfigure}{\thesection.\arabic{figure}}
\renewcommand{\theHtable}{\thesection.\arabic{table}}
\renewcommand{\theHequation}{\thesection.\arabic{equation}}
\makeatother

\widetext
\section*{Appendix A: Scheme of measurement-only-circuit}
In this appendix, we explain some details of the MoC and practical methods to calculate
observables.
We also show how the system evolves in the MoC and reach a steady state.

\subsection{Calculation of TEE}
\begin{figure}[h]
\begin{center}
\includegraphics[width=10cm]{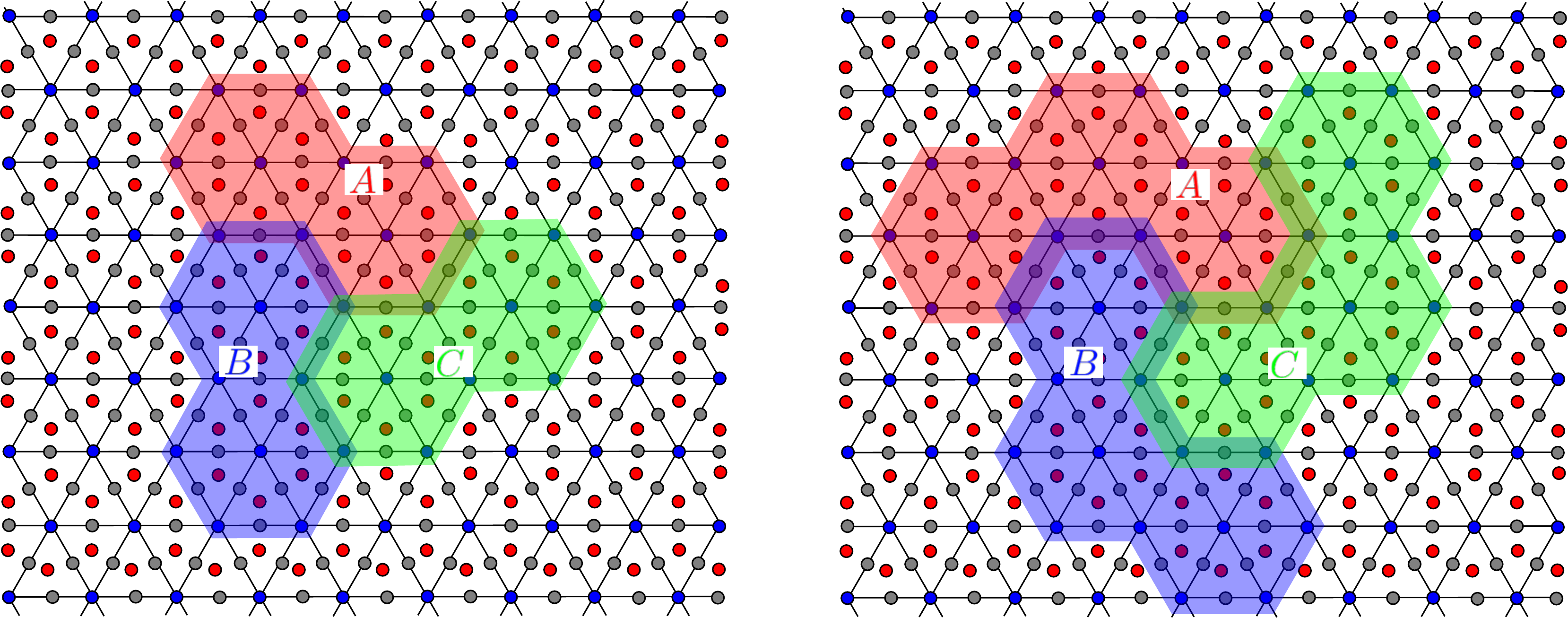}
\end{center}
\caption{Complexes of hexagons used for calculation of the TEE.
}
\label{Fig_hex}
\end{figure}

We first explain the methods to calculate the TEE.
The EE of subsystem $A$, $S_A$, is given by the number of the independent stabilizers $\{Z_e,X_e,A_s,B_p\}$
that connect the subsystem $A$ and its complement $A^c$\cite{Fattal2004}.
In order to calculate the TEE accurately, the value of $S_A$ has to satisfy 
\be
S_A=-L_A+\gamma, 
\label{SAGa}
\ee
where $L_A$ is the length of the subsystem $A$'s boundary, and $\gamma$
is a universal constant such as $\gamma=-1(0)$ for the TC (trivial state) \cite{Wen_text}.
Under the condition that the above equality is satisfied by numerics, the target TEE, $S_{\cal T}$,
is obtained by using the combination of the EE such as, 
\be
S_{\cal T}=S_A+S_B+S_C-S_{AB}-S_{BC}-S_{CA}+S_{ABC}.
\nonumber
\ee
We numerically examined if Eq.~(\ref{SAGa}) is satisfied for various shapes of subsystem in the TC and 
trivial states, and found that hexagon complexes are most suitable shapes of subsystem. 
We verified that fluctuations of data of $S_{\cal T}$ are sufficiently suppressed for the states in the MoC 
after a long period.
Some examples of subsystems used in the calculation are shown in Fig.~\ref{Fig_hex}.

\subsection{Time evolution of states in MoC and emergence of steady states}

\begin{figure}[t]
\begin{center}
\includegraphics[width=6cm]{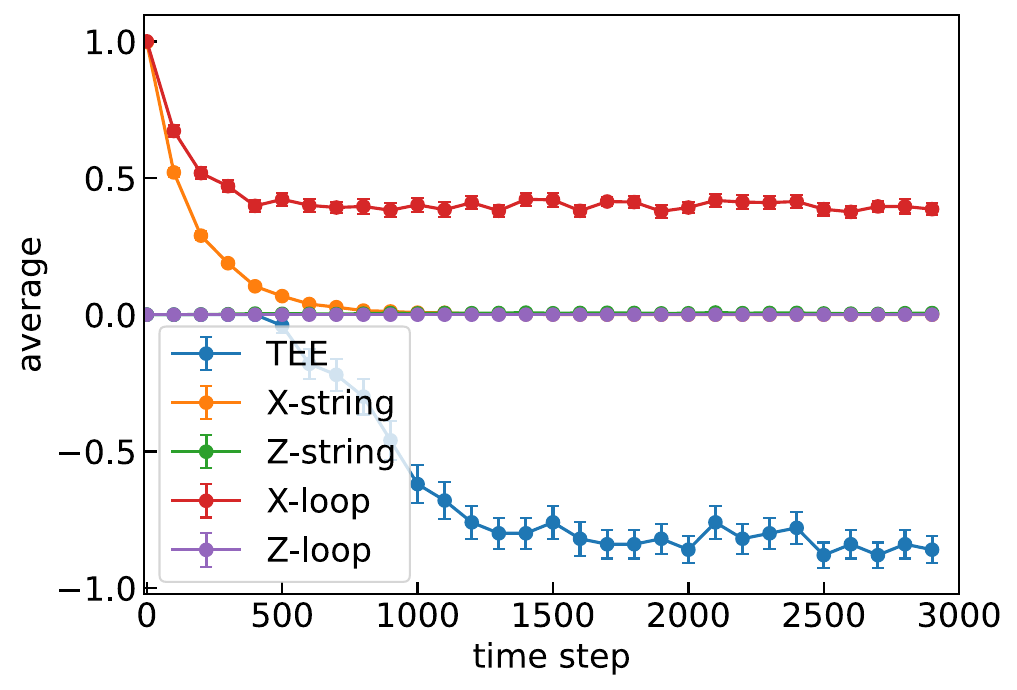}
\includegraphics[width=6cm]{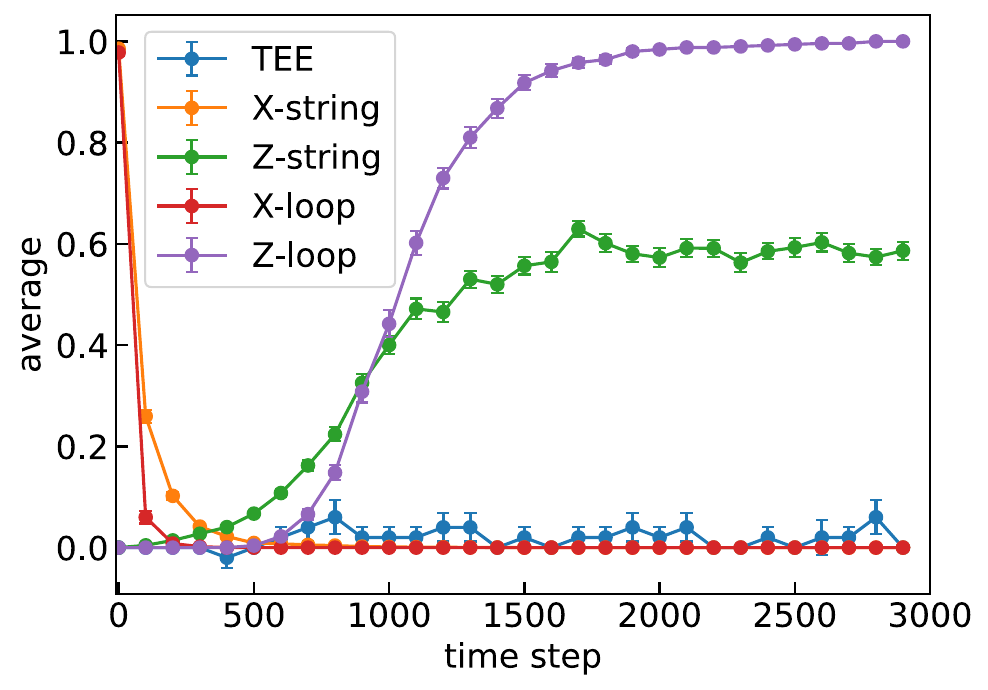}
\end{center}
\caption{Time evolution of states under the MoC.
Left: $p_x=0$, $p_z=0.1$ and $p_g=0.9$.
Right: $p_x=0, p_z=0.5$ and $p_g=0.5$.
All observables are getting steady after a sufficiently long period.
}\label{Fig_timestep_A2}
\end{figure}

In this subsection, we show how a state evolve in the MoC and verify that the period of the MoC
(the number of time steps) is large enough for emergence of a steady state. 
The order of the time step $\mathcal{O}(10^3)$ is sufficient for observing the properties of the steady states.
Typical examples are shown in Fig.~\ref{Fig_timestep_A2}. 

\section*{Appendix B: Details of finite-size scaling and critical exponents}

\begin{table*}[htbp]
\centering
    \begin{tabular}{|c|c|c|}    
    \multicolumn{3}{c}{Table 2A}    \\\hline
    \multicolumn{3}{|c|}{$p_x+p_g=1$}     \\ \hline
    \multicolumn{1}{|c|}{observable}   &  $p_g$  & \multicolumn{1}{c|}{$\nu$}  \\ \hline
    TEE$^\ast$   & 0.846 $\pm$ 0.016& 1.69 $\pm$ 0.26\\
    TEE & 0.851 $\pm$ 0.002& 1.67 $\pm$ 0.09\\
    TEE(1$\sim$6)& 0.846 $\pm$ 0.019& 1.69 $\pm$ 0.57\\ \hline
  \end{tabular}
   \begin{tabular}{|c|c|c|}    
    \multicolumn{3}{c}{Table 2B}    \\\hline
    \multicolumn{3}{|c|}{$p_z+p_g=1$}     \\ \hline
    \multicolumn{1}{|c|}{observable}   &  $p_g$  & \multicolumn{1}{c|}{$\nu$}  \\ \hline
    TEE$^\ast$   & 0.876 $\pm$ 0.048& 1.74 $\pm$ 0.29\\
    TEE  & 0.882 $\pm$ 0.069& 1.62 $\pm$ 0.80\\
    TEE(1$\sim$ 6)& 0.884 $\pm$ 0.050& 1.71 $\pm$ 0.38\\ \hline
  \end{tabular}
\vspace{0.5 cm}
\caption{Table 2A: Values of critical point and critical exponent estimated by
the various FSS scheme for the TEE obtained in the MoC with $p_x+p_g=1$.
TEE* is the value obtained by the bootstrap methods.
Table 2B: Values of critical point and critical exponent estimated by the various FSS 
scheme for the TEE obtained in the MoC with $p_z+p_g=1$.
TEE* is the value obtained by the bootstrap methods.
Data TEE(1$\sim 6$) is obtained for subsystems composed of $1\sim 6$ hexagons.
See Fig.~\ref{Fig_hex}.
The others are data of $2\sim 6$ hexagons. }
\label{Tab2}
\end{table*}

In this appendix, we show details of the FSS used to obtain the transition point and critical
exponents observed by the TEE.
We mostly study the variance of the TEE and its FSS by using pyfssa.
To extract the transition point and critical exponents, we examined two kinds of analysis to 
verify the robustness of the obtained numerical values.
We summarize the results in Table \ref{Tab2}.
The mainly employed method is the bootstrap analysis using $500$ trajectory samples of the MoC,
with which $1000$ pseudo-ensembles containing $500$ TEE data are constructed by randomly choosing
the TEE data with permitting multiple choosing.
Then, the mean values of the critical $p_g$ and the critical exponents are obtained.
The other, simpler method calculates the variance of the 500 TEE samples and estimates its standard error to be one half of 
the standard error of the sample mean (SEM).
The bootstrap-based estimation provides a statistically robust evaluation of fluctuations without assuming any specific distribution 
and is therefore adopted as the primary method.
For reference, one half of the SEM was also used as an approximate measure of the standard error of the variance to maintain consistent scaling of the uncertainties across different system sizes and sampling conditions.
The bootstrap- and SEM-based estimates were found to be of comparable magnitude for the present normalized dataset, 
demonstrating that the SEM-based approximation is valid and does not affect the overall evaluation.
Value of the TEE is also slightly depends on which sample data is used, e.g., the number of hexagon 
complex.
In Table \ref{Tab2}, we summarize estimated values of the TEE by various methods and find that 
the deviations are small.

\section*{Appendix C: Connection to Fredenhagen-Marcu operator}
\setcounter{subsection}{0}

\begin{figure}[t]
\begin{center}
\includegraphics[width=8.5cm]{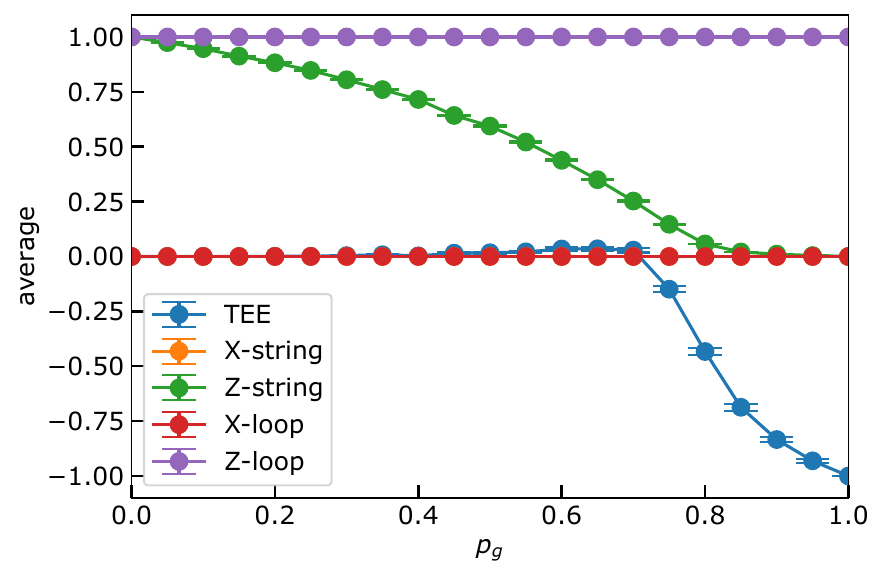}
\end{center}
\caption{Behavior of various observables from the Higgs to TC regimes with
$p_z+p_g=1$.
The initial state is the $Z$-product state.
The expectation value of the $Z$-loop is unity for the whole parameter region, 
whereas that of the $Z$-string is a decreasing function of $p_g$, and vanishes in the TC regime.
This result indicates that the emergent TC is an eigenstate of the $Z$-loop (the logical-$Z$) operator.
For any value of $p_g$, the expectation value of contractible $Z$-loops is unity as the stabilizer group is 
composed of $\{Z_e, B_p,A_s\}$, and therefore, the expectation value of string operators is closely related 
to the FMO.
}
\label{Fig_Higgs_TC}
\end{figure}

It is beneficial to see the connection between the order parameters used in the present study and 
the Fredenhagen-Marcu operator (FMO) \cite{Fredenhagen1983,PhysRevLett.56.223}.
The FMO provides a useful order parameter discriminating the Higgs phase and defined as follows,
\be
C_Z={\langle\Psi| \prod_{e\in L_{1/2}} Z_e |\Psi\rangle \over \sqrt{\langle\Psi| \prod_{e\in L}Z_e|\Psi\rangle}},
\label{FMO}
\ee
where $L$ is a loop whose length is twice of the length of the string $L_{1/2}$.
Denominator of $C_Z$ compensates for the perimeter-law decay of the numerator and 
$C_Z$ has a finite value in the Higgs phase, whereas it vanishes in the Coulomb phase for $L_{1/2}\to \infty$.

For the system under the present study, some combination of the order parameters has a close relation to the FMO.
To see this, let us consider the transition from the Higgs phase to the TC with the logical $Z$-operator, in which
the non-contractible $Z$-loop operators work as a logical qubit and have a finite expectation value.
The TC is a specific phase of the gauge theory, meaning that the string tension of $Z$ and $X$ loops vanishes.
As a result, the logical $Z$-operator (a non-contractible $Z$-loop operator) has the unity expectation value
regardless of its length in the TC state under consideration.
This fact obviously indicates that an arbitrary contractible $Z$-loop also has the unity expectation value,
as it is easily verified in the genuine TC.

In Fig.~\ref{Fig_Higgs_TC}, we show the string and loop order parameters for the regime from the Higgs to the TC.
As we explained in the above, the non-contractible $Z$-loop operators have unity expectation value
in both the Higgs and TC phases.
On the other hand, the expectation value of the $Z$-string operator decreases as increasing $p_g$
and vanishes in the TC regime.
Therefore, we can observe FMO-like behavior of the state from the string and loop $Z$-operators.
Although the above observation is obtained for the case with $p_x=0$, it is plausible to expect a similar behavior of the order parameters for small but finite $p_x$'s.
On the other hand, for a moderate value of $p_x$, the FMO is a better quantity to investigate the phase diagram.

\setcounter{subsection}{0}
\section*{Appendix D: Additional numerical data}
In this appendix, we show additional numerical data that are complements to the main text.

\subsection{TEE for the region $p_x\neq 0, p_z\neq 0$}
In the main text, we mostly showed the numerical results for $p_x=0$ or $p_z=0$.
This subsection displays the data of the TEE, string and loop operators for non-vanishing $p_x$ and $p_z$, 
where some competition between the stabilizers $\{A_s,B_p,Z_e,X_e\}$ is enhanced, and it is interesting 
to see how the observables behavior there.

\begin{figure}[t]
\begin{center}
\includegraphics[width=12cm]{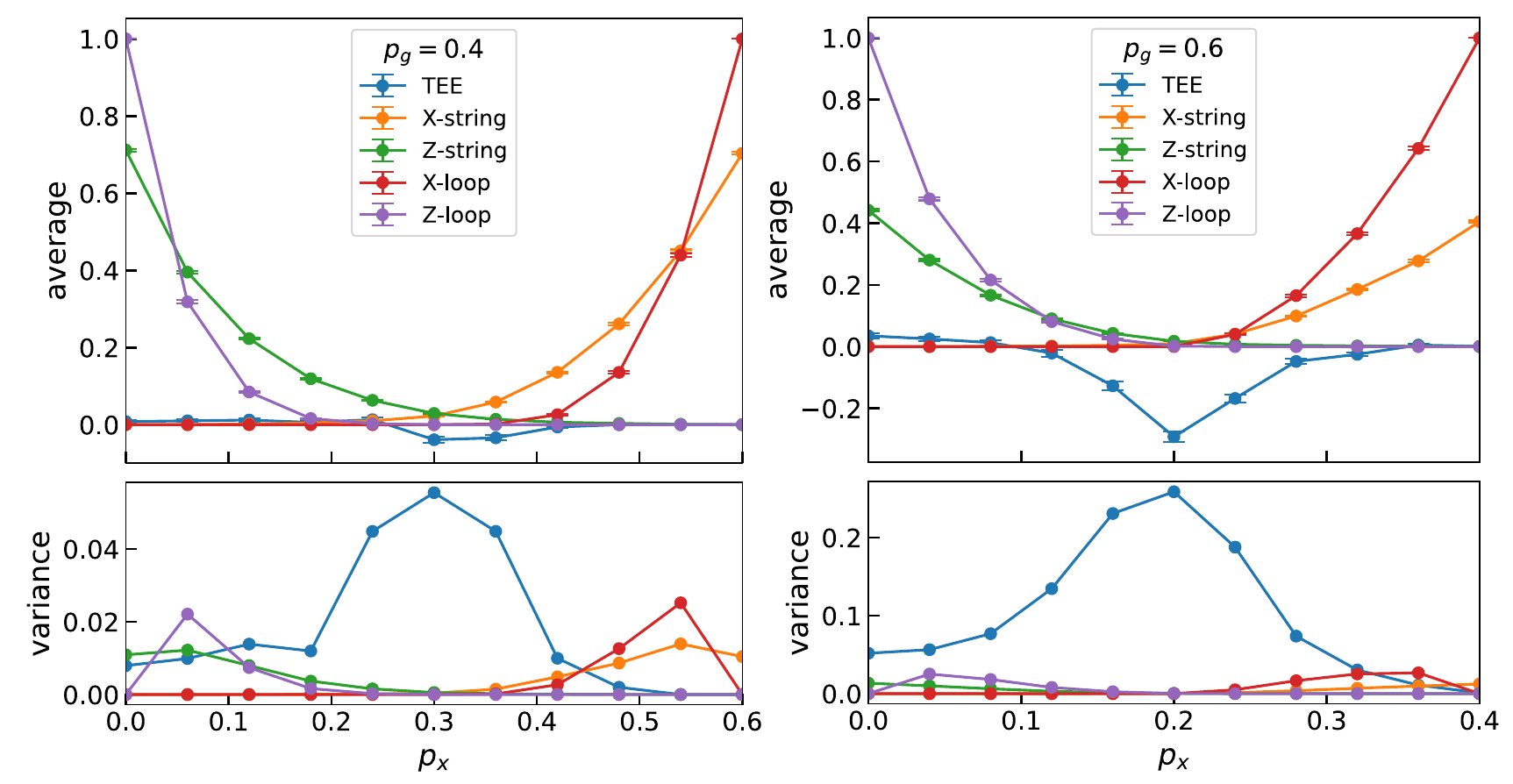}
\end{center}
\caption{Calculation of the TEE, Wilson and 't Hooft strings and loops for $p_g=0.4$ and
$p_g=0.6$ with keeping $p_x+p_z=1-p_g$.
In the both cases, the strings and loops exhibit quite different behavior in the confinement
and Higgs regimes.
For the states with $p_g=0.4$, the TEE slightly decreases for $p_x\sim p_z$.
On the other hand for $p_g=0.6$, the TEE decreases strongly and exhibits a cusp at $p_x=p_z$
showing a possible genuine phase transition.
Furthermore, the variance of the TEE exhibits a large peak for $p_g=0.6$ (compared with
the TEEs in Fig.~\ref{Fig4b}) supporting the existence of a phase transition.
On the other hand, for $p_g\leq 0.4$, at most a crossover connects the Higgs and confinement regimes.
}
\label{Fig_pgfixed}
\end{figure}

In the numerical study, we fixed the value of $p_g$ with keeping $p_x+p_z=p_g$, and calculated
the observables.
Numerical results are displayed in Fig.~\ref{Fig_pgfixed}.
For the both cases $p_g=0.4$ and $p_g=0.6$, the strings and loops exhibit behaviors that discriminate the confinement and Higgs regimes.
On the other hand, the TEE for $p_g=0.4$ shows only small decrease for $p_x\sim p_z$,
whereas for $p_g=0.6$, the TEE decreases rapidly for $p_x\sim p_z$ and a cusp appears at $p_x=p_z$,
indicating the existence of a phase transition. 
Furthermore, the variance of the TEE exhibits a large peak for $p_g=0.6$ (compared with
the TEEs in Fig.~\ref{Fig4b}) supporting the existence of a phase transition.
On the other hand, for $p_g\leq 0.4$, at most a crossover connects the Higgs and confinement regimes.
Anyway, more detailed study on this critical regime is an interesting future issue.

\subsection{Strings and loops for $p_z+p_g=1$}

\begin{figure}[t]
\begin{center}
\vspace{0.5 cm}
\includegraphics[width=12.2cm]{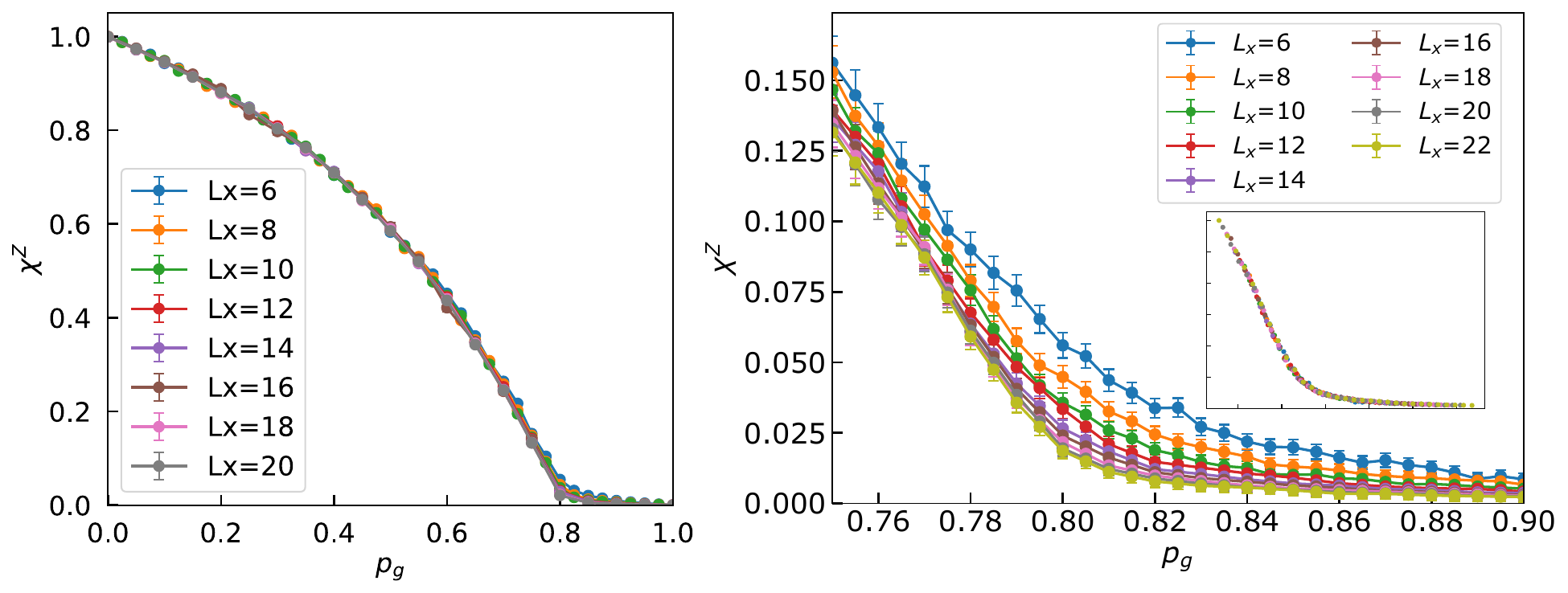}
\vspace{-0.5 cm}
\end{center}
\caption{Calculation of the expectation value of $Z$-string from the Higgs regime to the toric code.
In the inset, we observe the collapse of the date by finite-size scaling.
}
\label{Fig_Astringloop}
\end{figure}

\begin{figure*}
\includegraphics[width=17.5cm]{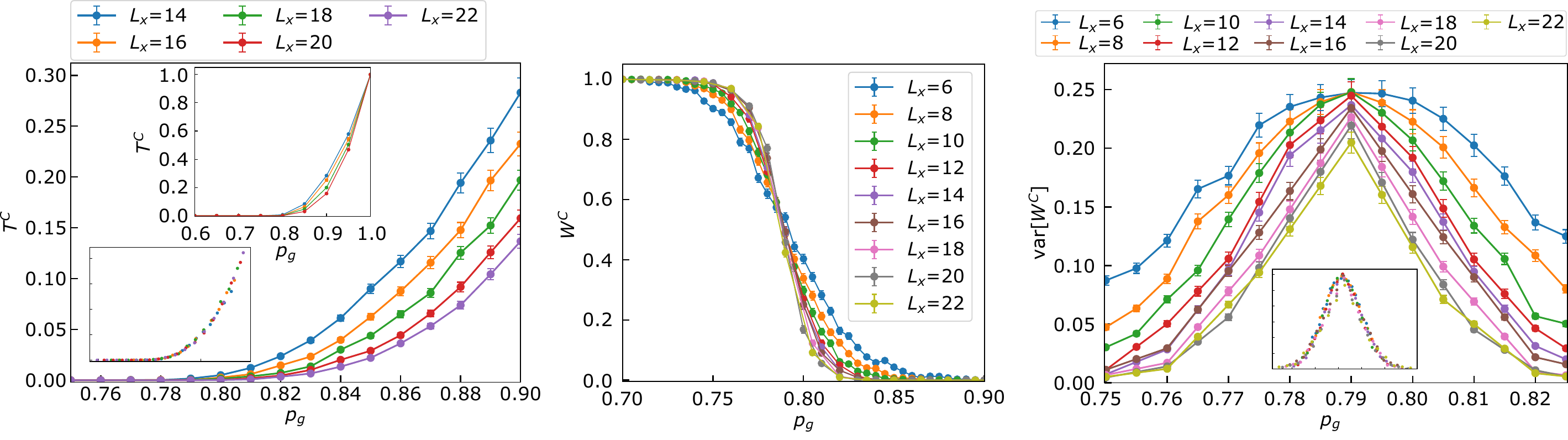}
\caption{
Left panel: Calculation of the expectation value of $X$-loop from the Higgs regime to the toric code.
In the inset, we observe the collapse of the date by finite-size scaling.
Center panel: Calculation of the expectation value of $Z$-loop from the Higgs regime to the toric code.
Right panel: Variance of $Z$-loop, and the inset shows collapse of data by the FSS.
}
\label{Fig_Aloop}
\end{figure*}

In the main text, we showed the behavior of strings and non-contractible loops in the MoC
with keeping $p_x+p_g=1$, in particular, how they behave in the critical regime.
In this subsection, we display the calculations of their expectation values for $p_z+p_g=1$.
As we explained in the main text, the measurement of $Z_e$ competes with the stabilizer $A_s$,
and therefore, we expect slightly different results from the case of $p_x+p_g=1$.
Results are displayed in Fig.~\ref{Fig_Astringloop} and Fig.~\ref{Fig_Aloop}.


\end{document}